\documentclass[aps,prb,twocolumn,showpacs,eqsecnum,superscriptaddress,floatfix]{revtex4-1}

\usepackage{graphicx}
\usepackage{epstopdf}
\usepackage{color}\usepackage{enumerate}
\usepackage{amsmath}\usepackage{amssymb}\usepackage{stmaryrd}\usepackage{wasysym}
\usepackage{multirow}
\usepackage{float}
\usepackage{longtable,booktabs}
\usepackage{hyperref}\usepackage{url}
\usepackage{subfigure}%
\usepackage{dsfont}

\begin{document}

\title{Twofold twist defect chains at criticality}

\author{Xiongjie Yu}
\affiliation{Department of Physics and Institute for Condensed Matter Theory, University of Illinois at Urbana-Champaign, 1110 West Green Street, Urbana, IL 61801-3080, USA}
\author{Xiao Chen}
\affiliation{Kavli Institute for Theoretical Physics, University of California at Santa Barbara, CA 93106, USA}
\affiliation{Department of Physics and Institute for Condensed Matter Theory, University of Illinois at Urbana-Champaign, 1110 West Green Street, Urbana, IL 61801-3080, USA}
\author{Abhishek Roy}
\affiliation{Institute of Theoretical Physics, University of Cologne, Zulpicher Strasse 77, D-50937 Cologne, Germany}
\author{Jeffrey C. Y. Teo}
\affiliation{Department of Physics, University of Virginia, Virginia 22904, USA}

\date{\today}

\begin{abstract}
The twofold twist defects in the $D(\mathbb{Z}_k)$ quantum double model (abelian topological phase) carry non-abelian fractional Majorana-like characteristics. We align these twist defects in a line and construct a one dimensional Hamiltonian which only includes the pairwise interaction. For the defect chain with even number of twist defects, it is equivalent to the $\mathbb{Z}_k$ clock model with periodic boundary condition (up to some phase factor for boundary term), while for odd number case, it maps to $\mathbb{Z}_k$ clock model with duality twisted boundary condition. At critical point, for both cases, the twist defect chain enjoys an additional translation symmetry, which corresponds to the Kramers-Wannier duality symmetry in the $\mathbb{Z}_k$ clock model and can be generated by a series of braiding operators for twist defects. We further numerically investigate the low energy excitation spectrum for $k=3,~4,~5$ and $6$. For even-defect chain, the critical points are the same as the $\mathbb{Z}_k$ clock conformal field theories (CFTs), while for odd-defect chain, when $k\neq 4$, the critical points correspond to orbifolding a $\mathbb{Z}_2$ symmetry of CFTs of the even-defect chain. For $k=4$ case, we numerically observe some similarity to the $\mathbb{Z}_4$ twist fields in $SU(2)_1/D_4$ orbifold CFT.

\end{abstract}

\maketitle

\section{Introduction}
Non-abelian anyons, such as Ising and Fibonacci anyons have non-abelian braiding statistics and can store quantum information non-locally.\cite{Kitaevchain, OgburnPreskill99, Preskilllecturenotes, FreedmanKitaevLarsenWang01, ChetanSimonSternFreedmanDasSarma} Such a state can be used as quantum memory and have promising application in topological quantum computing. These non-abelian anyons are expected to exist in a non-abelian fractional quantum Hall liquid.\cite{MooreRead, GreiterWenWilczek91, NayakWilczek96}

Recently, topological defects with non-abelian braiding statistics have been predicted in the abelian topological phases.\cite{Kitaev06, KitaevKong12, Bombin, YouWen, YouJianWen, TeoRoyXiao13long, teo2013braiding, khan2014, BarkeshliQi, BarkeshliJianQi, BarkeshliJianQi13, BarkeshliJianQi13long, Teotwistdefectreview}
These topological defects are present at the heterostructures and dislocations in some abelian topological states.\cite{HasanKane10,QiZhangreview11,FuKane08, AkhmerovNilssonBeenakker09, FuKanechargetransport09, LawLeeNg09, GoldhaberGordon12, BarkeshliJianQi, mongg2} They can carry (fractional) Majorana-like characteristics and are manifested as the twist defects in topological phases with global symmetries, such as Kitaev toric code, the Bombin-Martin color code and its $\mathbb{Z}_k$ generalization.\cite{Kitaev06, KitaevKong12, Bombin, BombinMartin06, YouWen, YouJianWen, TeoRoyXiao13long, teo2013braiding, khan2014, BarkeshliQi, BarkeshliJianQi, BarkeshliJianQi13, BarkeshliJianQi13long,BarkeshliBondersonChengWang14, Teo2015, Tarantino2016} There has been theoretical proposals for their realization in superconductor (SC)-- (anti)ferromagnet (FM) -- (fractional) topological insulator (TI) heterostructures, where Majorana zero modes or parafermions for the fractional case are bounded at the point defect interfaces.\cite{FuKane08,AkhmerovNilssonBeenakker09, FuKanechargetransport09,LawLeeNg09,FuKaneJosephsoncurrent09,Wilczek09,HasslerAkhmerovBeenakker11,Alicea12,GoldhaberGordon12, BarkeshliJianQi,ClarkeAliceaKirill,LindnerBergRefaelStern,MChen,Vaezi,mongg2,Teotwistdefectreview}

For example, in the Kitaev's toric code model~\cite{Kitaev97},
the twofold twist defect~\cite{Kitaev06,Bombin} that associates with the electric-magnetic duality symmetry changes the $\mathbb{Z}_2$ gauge charge ${\bf e}$ into the gauge flux ${\bf m}$, or vice versa, when the quasiparticle orbits around the defect (fig.~\ref{twist_defect}). Due to the non-local twisting structure, the topological defect carries a non-trivial quantum dimension $d=\sqrt{2}$, and the defect system can be physically mapped~\cite{KhanTeoVishveshwara15} on to the SC-FM-TI heterostructure~\cite{FuKane08} that supports Majorana zero mode.
In general, twist defects are extrinsic classical point defects in topological phases associating with a global anyonic symmetry $g$.\cite{khan2014,BarkeshliJianQi13,Teotwistdefectreview} The twist defect permutes the anyon labels of orbiting quasiparticles and acts as fluxes of anyonic symmetry.
They are non-abelian objects and their fusion and braiding properties can be systematically described by a defect fusion category or a $G$-crossed tensor category.\cite{TeoRoyXiao13long,BarkeshliBondersonChengWang14, Teo2015, Tarantino2016, Teotwistdefectreview}

In this paper, we will consider one dimensional chains of twist defects and study the critical point of these chain modes. These twist defects are embedded in the background of a $D(\mathbb{Z}_k)$ quantum double model, which is a $\mathbb{Z}_k$ generalization of Kitaev $\mathbb{Z}_2$ toric code model, and can also be understood as a discrete $\mathbb{Z}_k$ gauge theory in its deconfined phase.\cite{BaisDrielPropitius92,Propitius-1995, PropitiusBais96, Kitaev97,Preskilllecturenotes,FreedmanLarsenWang00,Mochon04,Bais-2007} The twist defects here are non-abelian defects and carry zero modes of $\mathbb{Z}_k$ parafermions.\cite{Baxterbook, Fradkin1980, Fendley12, Mong2014} These are twofold defects in the sense that the corresponding anyonic symmetry operation is of order two, and that a pair of defects associates to a $k$-dimensional Hilbert space. We introduce pairwise interaction between twist defects and construct the defect chain Hamiltonian. Similar ideas have been used before to construct the non-abelian anyonic chain models and study the phase diagram in them.\cite{FeiguinTrebstLudwigTroyerKitaevWangFreedman07, Poilblanc2011, GilsArdonneTrebstHuseLudwigTroyerWang13, Pfeifer2012, Li2015}

In our model, the pairwise interaction can be represented by the Wilson loop operator around the neighboring twist defects\cite{TeoRoyXiao13long} that separates the $k$ quantum states. Based on the algebra of Wilson loop operators, we will show that the twist defect chain model with periodic boundary condition can be mapped to the $\mathbb{Z}_k$ clock model with various boundary conditions. For even number of twist defects, the corresponding $\mathbb{Z}_k$ clock model has periodic boundary condition (up to a phase for the boundary term), while for odd number of twist defects, after mapping to $\mathbb{Z}_k$ clock model, this requires the introduction of a new type of boundary condition. This boundary term was studied in the $k$-state Potts model ($k\leq 4$) using the language of Temperley-Lieb algebra and was called a duality twisted boundary condition.\cite{Schutz1993} In both even and odd cases, we will show that at critical point, the twist defect chain model preserves translational symmetry, which is identical to the Kramers-Wannier duality symmetry in the clock model setting. We will show that the translational symmetry operator has a simple physics interpretation and can be understood as a product of braiding operators which exchanges the positions of neighboring twist defects.

The critical point of the lattice models in $1+1$ dimensions can be described by rational conformal field theory (CFT), which has finite number of primary fields.\cite{Moore1990,bigyellowbook} For an even chain, it corresponds to the $\mathbb{Z}_k$ clock model at criticality and the structure of the CFT is well known.\cite{Jose1977, Elitzur1979} The simplest example is $k=2$ case, which is the critical transverse field Ising model with central charge $c=1/2$. For an odd chain, the underlying CFT is not well-studied in the literature in general when $k\neq2$. For $k=2$, the odd chain corresponds to the transverse field Ising model with a duality twisted boundary condition. This model can be mapped to a free fermion chain under Jordan-Wigner transformation and can been calculated analytically. Under this twisted boundary condition, the partition function takes a non-diagonal form in terms of characters, in which the holomorphic or anti-holomorphic character (depending on the phase of the boundary term) has conformal dimension $h=1/16$.\cite{Grimm2002}

For $k>2$ cases with duality twisted boundary condition, the model is not interaction-free anymore and therefore an analytical result is absent. In the present article, we will numerically study the energy spectra of these models for $k=3,~4,~5,~6$ at criticality and extract the conformal scaling dimensions for the primary fields of the underlying CFT. Based on these result, we will demonstrate that when $k\neq 4$, as a CFT, the odd-chain models can simply be related to the even-chain models with some additional twofold twist field operators. However, special care is needed for $k=4$, where we find that new excitations are consistent with some fourfold twist field operators in the $SU(2)_1/D_4$ CFT. Such kind of CFT is the so-called orbifold CFT and has been extensively studied in the literature.\cite{DijkgraafVafaVerlindeVerlinde99, Ginsparg88, Ginsparg1988, ChenRoyTeoRyu17} We summarize the main results in Table~\ref{summary_even_odd}.

\begin{table}[htbp]
\centering
\begin{tabular}{c|c|c|c}
$k$ & coupling & Even-chain & Odd-chain \\\hline
$3$ & F & three-state Potts & $\mathcal{M}(5,6)$\\
$3$ & AF & $U(1)_3$ & $U(1)_3/\mathbb{Z}_2$\\
\hline
$4$ & F/AF & $U(1)_2/\mathbb{Z}_2$ & $SU(2)_1/D_4$ ? \\
\hline
$k>4$ & F/AF & $U(1)_k$ & $U(1)_k/\mathbb{Z}_2$\\
\end{tabular}
\caption{The underlying CFT for chains with even number of defects (the third column) and the underlying CFT for chains with odd number of defects (the fourth column). $U(1)_k$ refers to the K-matrix $K=2k$ in the boson Lagrangian density $\mathcal{L}=(K/2\pi)\partial_t\phi\partial_x\phi$, and the details can be found in Appendix~\ref{U_1_orb}. For the odd-chain with $k=4$, the numerical results show certain similarities with the $SU(2)_1/D_4$ orbifold CFT.}
\label{summary_even_odd}
\end{table}

The rest of this paper is as follows. In Sec.~\ref{model}, we first briefly review the twist defect in topological phase and then we construct the Wilson loop  Hamiltonian with  even number and odd number of twist defects. We also discuss the translational symmetry in both cases. In Sec.~\ref{numerics}, we first explain our numerical method and then calculate the primary fields for even number case. We further study the odd number case and extract the conformal dimension for the twist field operator. We summarize and conclude in Sec.~\ref{conclusion}. The appendices are devoted to details of the calculations and techniques used in this paper.

\section{Twist defect chain model}
\label{model}
\subsection{Review of twist defect}
The $D(\mathbb{Z}_k)$ quantum double model in $2+1$ dimensions is the $\mathbb{Z}_k$ lattice gauge theory in the deconfined limit and is an abelian topological phase.  It has two fundamental excitations the gauge charge $\textbf{e}=(1,0)$ and the gauge flux $\textbf{m}=(0,1)$ and all the $k^2$ quasi-particle excitations can be written as $\textbf{a}={\bf e}^s{\bf m}^t$ with $0\leq s,t\leq k$.  The braiding phase between $\textbf{e}$ and $\textbf{m}$ is $e^{2\pi i/k}$. For the $\mathbb{Z}_2$ case, the toric code is related to the s-wave superconductor with a deconfined $\mathbb{Z}_2$ fermion parity symmetry by identifying ${\bf m}$ with the the $hc/2e$ flux vortex, ${\bf em}$ with the BdG fermion, and ${\bf e}$ with an excited vortex.\cite{KhanTeoVishveshwara15, KhanTeoHughesVishveshwara16,Teo2015}

\begin{figure}[h]
	\centering
	\includegraphics[scale=.4]{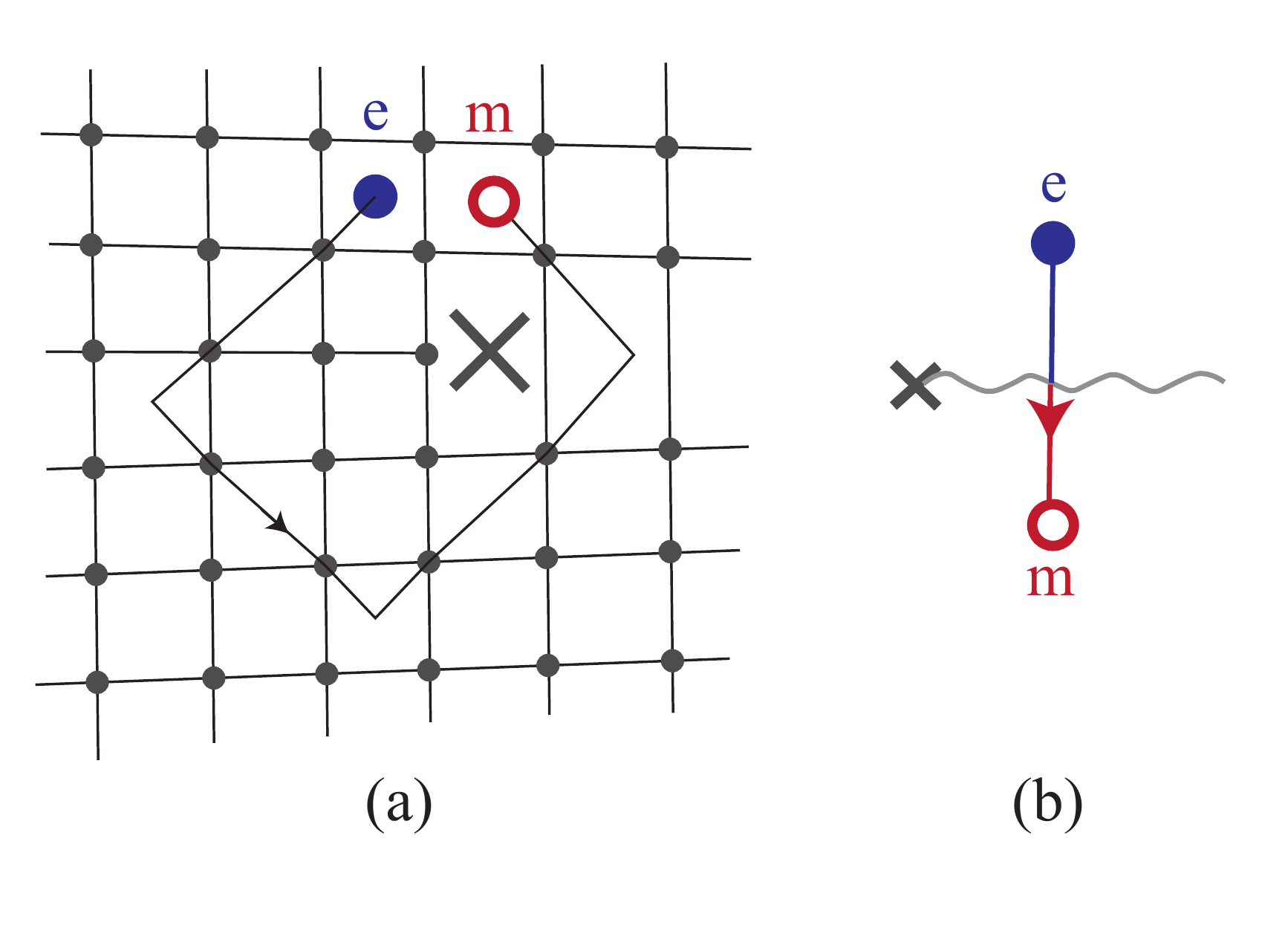}
	\caption{ (a) Twofold twist defect realized as a lattice dislocation in Wen's plaquette model. Anyon type flips between ${\bf e}$ and ${\bf m}$ when orbiting around the twist defect. (b) An arbitrary artificial branch cut signaling the anyon type flip of a crossing quasiparticle. }
	\label{twist_defect}
\end{figure}

The $D(\mathbb{Z}_k)$ quantum double model  has the global duality symmetry operation, which will interchange $\textbf{e}$ and $\textbf{m}$ excitations. As shown in Fig.\ref{twist_defect}, in the lattice model (Wen's plaquette model \cite{Wenplaquettemodel}), the global duality symmetry operation corresponds to the half lattice translation and will interchange the $\textbf{e}$ and $\textbf{m}$ particles which live on blue and red plaquettes respectively. In this sense, the topological phase weakly breaks the global duality/half lattice translation symmetry. The duality symmetry can be partially restored by introducing twofold twist defect, which is the dislocation on the lattice model as shown in Fig.\ref{twist_defect}.\cite{Kitaev06, Bombin, BarkeshliJianQi, BarkeshliQi, YouWen, YouJianWen, KitaevKong12, khan2014, BarkeshliBondersonChengWang14, TeoRoyXiao13long, teo2013braiding} The twist defect can be pictorially represented by a cross attached with a branch cut. After crossing the branch cut, the $\textbf{e}$ particle and the $\textbf{m}$ particle will be interchanged. The twist defects are semi-classical non-abelian defects and each pair of them can form a k-level system. When $k=2$, the twist defect corresponds to the more familiar Majorana zero mode.\cite{HasanKane10,QiZhangreview11} In our previous work, we systematically studied the fusion rule and F-symbols for basis transformation in a multi-defect system.\cite{TeoRoyXiao13long,teo2013braiding} The fusion between the twofold defects and abelian anyon are given by
\begin{eqnarray}
\nonumber &&\textbf{a}\times\sigma_{\lambda}=\sigma_{\lambda+s+t},\\
&&\sigma_{\lambda_2}\times\sigma_{\lambda_1}=\sum_{\lambda_1+\lambda_2=s+t}\textbf{a}
\label{fusion}
\end{eqnarray}
where $\lambda$ runs from $0$ to $k-1$ mod $k$ and is the species label for the twist defect and $\textbf{a}=\textbf{e}^s\textbf{m}^t$ is the abelian anyon.
The unitary braiding operator for twist defects projectively represents the sphere braid group.


\subsection{Wilson loop Hamiltonian}

\begin{figure}[h]
	\centering
	\includegraphics[scale=.34]{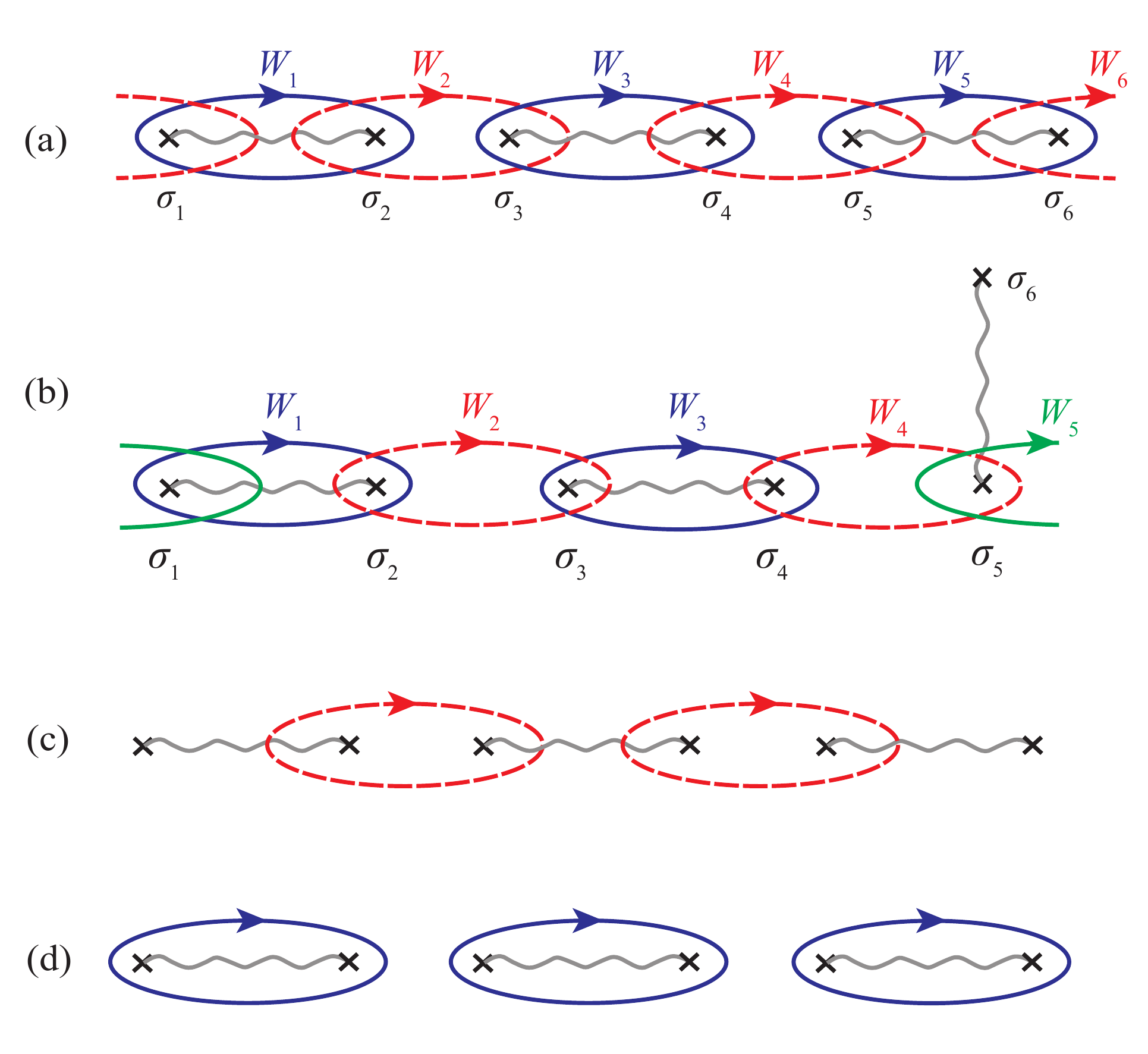}
	\caption{ (a) Twist defect chain with six (even number) twofold defects. The twist defect at site $a$ is labelled as $\sigma_a$. $\mathcal{W}_a$ is the Wilson loop operator around $\sigma_a$ and $\sigma_{a+1}$. In terms of $\mathbb{Z}_k$ clock variable, $\mathcal{W}_{2j-1}$ (solid blue curve) corresponds to the term $\tau_j$, while $\mathcal{W}_{2j}$ (dashed red curve) represents the term $\sigma^{\dag}_j\sigma_{j+1}$. The Wilson loop Hamiltonian has periodic boundary condition here. (b) Twist defect chain with five (odd number) twofold defects, with $\sigma_6$ is pulled away far from the defect chain. The branch cut connecting $\sigma_5$ is left behind. The boundary term $\mathcal{W}_5$ (with green solid curve) is the boundary term and corresponds to the duality twisted boundary condition in the $\mathbb{Z}_k$ clock model. (c) The Hamiltonian of Wilson loop operator only at even site. Here we choose open boundary condition and therefore $\sigma_1$ and $\sigma_6$ do not show up in the Hamiltonian. In terms of $\mathbb{Z}_k$ clock variable, this Hamiltonian is the same as Eq.\eqref{H_Wilson} with $J_2=0$ and no boundary term. (d) The Hamiltonian with Wilson loop operator only at odd site. It also corresponds to Eq.\eqref{H_Wilson} with $J_1=0$.}
	\label{defect_chain_dia}
\end{figure}

In this paper, we will use the bare twofold defect in the $D(\mathbb{Z}_k)$ quantum double model to construct some one dimensional chain models and in particular, we will focus on their critical behavior. The setup is like this, we first create $M$ bare twist defects in the background of $D(\mathbb{Z}_k)$ quantum double model and align them in a line. Each twist defect is attached with a branch cut and two of them can pair up by gluing the branch cut together. This pairing procedure is arbitrary and for simplicity, we connect $\sigma_{2j+1}$ and $\sigma_{2j+2}$ by the branch cut as shown in Fig.\ref{defect_chain_dia}. For the even case with $M=2N$, all twist defects can pair up and there are no branch cut left. While for the odd case with $M=2N-1$, the last twist defect $\sigma_{2N-1}$ cannot find twist defect to pair up with and has a dangling branch cut left behind. The quantum dimension for the total Hilbert space is $k^{M/2}$. For convenience, we will denote the bare twofold defect at site $a$ as $\sigma_{a}$. We use the Wilson loop operators to construct a one dimensional Hamiltonian (Fig.\ref{defect_chain_dia} (a))
\begin{equation}
H=-\sum_{a=1}^{M}J_a(\mathcal{W}_a+\mathcal{W}^{\dag}_a)
\label{Hamiltonian}
\end{equation}
where each Wilson loop operator is generated by dragging an $\textbf{e}$ particle around two neighboring twist defects. The Wilson operator is also dyon tunneling (fermion tunneling for the $\mathbb{Z}_2$ case) between neighboring twist defects. When $1\leq a<M$, $\mathcal{W}_{a}$ is the Wilson loop circling around $\sigma_{a}$ and $\sigma_{a+1}$, and the boundary term $\mathcal{W}_{M}$ is the Wilson loop circling around $\sigma_{M}$ and $\sigma_{1}$. For $\mathbb{Z}_2$ case, this is just the Majorana chain.\cite{Kitaevchain} We are interested in constructing the translational invariant model for the bare twist defects. This requires that  in Eq.\eqref{Hamiltonian}, all the $J_a$ are equal up to a phase. The detail for this phase will be explained later in Sec.\ref{even_num} and Sec.\ref{odd_num}.

According to the fusion rule in Eq.\eqref{fusion}, the two neighboring twist defects can fuse into an abelian anyon, the Wilson loop operator around these two twist defects can be used to detect the fusion channel of the two twist defects. The Wilson loop operator has eigenvalue equal to $e^{2\pi n i/k}$ with $n=0,...,k-1$ and satisfies $\mathcal{W}_a^k=1$. The commutation relationship between different Wilson loop operators are determined by the intersection between them,
\begin{eqnarray}
\nonumber &&[\mathcal{W}_{a},\mathcal{W}_{b}]=0,\quad \mbox{when}\  |a-b|>1\\
\nonumber &&\mathcal{W}_{a}\mathcal{W}_{a+1}=\omega\mathcal{W}_{a+1}\mathcal{W}_{a}\\
&&\mathcal{W}_{a+1}^{\dag}\mathcal{W}_{a}=\omega\mathcal{W}_{a}\mathcal{W}_{a+1}^{\dag}
\label{intersection}
\end{eqnarray}
where $\omega=e^{2\pi i/k}$.

This model in Eq.\eqref{Hamiltonian} is invariant under translation symmetry operator and under this symmetry, $\mathcal{T}\mathcal{W}_{a}\mathcal{T}^{-1}=\mathcal{W}_{a+1}$ . $\mathcal{T}$ operator can be realized by moving the last defect $\sigma_{M}$ all the way back to the first and can be represented by a sequence of braiding operators
\begin{equation}
\mathcal{T}=B_{1}B_{2}...B_{M-1}
\label{translation}
\end{equation}
where the braiding operator $B_{i}$ denotes a counter-clockwise permutation of a pair of adjacent defects at position $i$ and $i+1$.\cite{TeoRoyXiao13long, teo2013braiding,LindnerBergRefaelStern} We will show the translational symmetry operator $\mathcal{T}$ is the Kramers-Wannier duality symmetry and guarantees that the model is at the critical point.\cite{Fradkinbook}

According to the definition of Wilson loop algebra in Eq.\eqref{intersection}, the Wilson loop operator can be denoted as a $\mathbb{Z}_k$ clock variable with $\mathcal{W}_{2j-1}=\sigma_j$ ($\sigma$ here does not mean the twist defect) and $\mathcal{W}_{2j}=\tau_j\tau_{j+1}^{\dag}$, where $\tau$ and $\sigma$ are both $k$-dimensional matrices
\begin{equation}
\sigma=\begin{pmatrix}1&0&\cdots&0\\0&\omega&\cdots&0\\ \vdots&\vdots&\ddots &\vdots\\0&0&\cdots&\omega^{k-1}\end{pmatrix},\quad \tau=\begin{pmatrix}0&\cdots&0&1\\ 1&\cdots &0&0\\ \vdots&\ddots&\vdots&\vdots\\0&\cdots&1&0\end{pmatrix}
\end{equation}
with $\omega=e^{2\pi i/k}$. $\sigma$ and $\tau$ satisfy $\sigma^k=1$, $\tau^k=1$ and $\sigma\tau=\omega\tau\sigma$. The $\sigma$ operator here is a measurement of the quantum state (or $\mathbb{Z}_k$ parafermion parity) associates to the defect pair joined by a branch cut. The $\tau$ operator is a parafermion parity flip and the Wilson operator $W_{2j}$ flips the $\mathbb{Z}_k$ parity of the two pairs of defects next to it. The Hamiltonian in terms of $\mathbb{Z}_k$ clock variables takes this form (up to some boundary term $H_B$),
\begin{equation}
H=-J\sum_{j}(\sigma_j+\tau_j^{\dag}\tau_{j+1}+h.c.) + H_B
\end{equation}

This is quantum $\mathbb{Z}_k$ clock model at critical point and there has been a long history of studying this model.\cite{Baxterbook, Fradkin1980, Fendley12, Li2015} Notice that there is a subtle difference between twist defect chain and $\mathbb{Z}_k$ clock model. The single-site translation operation in $\mathbb{Z}_k$ model actually corresponds to two-site translation operator $\mathcal{T}^2$ in the twist defect chain model. This indicates that in the $\mathbb{Z}_k$ clock model, the unit cell is doubled. The  $\mathcal{T}$ symmetry operator corresponds to the famous Kramers-Wannier duality symmetry operator in  $\mathbb{Z}_k$ clock model,\cite{Fradkin1980}
\begin{equation}
D\sigma^{\dag}_a\sigma_{a+1}D^{-1}=\tau_{a+1},\quad D\tau_{a}D^{-1}=\sigma^{\dag}_{a}\sigma_{a+1}
\label{KW_duality}
\end{equation}

Since the unit cell is doubled, the ordinary quantum $\mathbb{Z}_k$ clock model always corresponds to the twist defect chain with even number of twist defects $2N$. For the defect chain with odd number of twist defect $2N-1$, if it is written in terms of $\mathbb{Z}_k$ clock model, the boundary term will be modified. We will explain these two different cases in the following subsections.

\subsection{Wilson loop Hamiltonian with $2N$ twist defects}
\label{even_num}

For the Wilson loop Hamiltonian with $2N$ number of twist defects and with periodic boundary condition shown in Eq.\eqref{Hamiltonian}, the corresponding $\mathbb{Z}_k$ clock model also has periodic boundary condition.
The Hamiltonian in terms of $\mathbb{Z}_k$ variables takes this form
\begin{equation}
H=-J\sum_{j=1}^{N-1}(\sigma_j+\tau_j^{\dag}\tau_{j+1}+h.c.)-J(\sigma_{N}+\tau_1^{\dag}\tau_N+h.c.)
\label{even_chain}
\end{equation}

This is the $\mathbb{Z}_k$ clock model at the critical point. If not at the critical point, the translational invariant $\mathbb{Z}_k$ clock model with periodic boundary condition is
\begin{eqnarray}
\nonumber H&=&-J_2\sum_{j=1}^{N} (\sigma_j+\sigma_j^{\dag})-J_1\sum_{j=1}^{N-1} (\tau_j^{\dag}\tau_{j+1}+\tau_j\tau^{\dag}_{j+1})\\
\nonumber &&-J_1(\tau_N^{\dag}\tau_{1}+\tau_1\tau^{\dag}_{N})\\
\nonumber &=&-J_2\sum_{j=1}^N  (\mathcal{W}_{2j-1}+\mathcal{W}_{2j-1}^{\dag})-J_1\sum_{j=1}^{N} (\mathcal{W}_{2j}+\mathcal{W}_{2j}^{\dag})\\
\label{H_Wilson}
\end{eqnarray}

The above model has a global $\mathbb{Z}_k$ symmetry and therefore we can define a global $\mathbb{Z}_k$ charge operator
\begin{align}
Q=\prod_{j=1}^{N}\sigma_j
\label{z_k_charge}
\end{align}
We briefly explain the phases for Eq.\eqref{H_Wilson} here. For the $\mathbb{Z}_k$ clock model, there are two limits, one is $|J_2|<<|J_1|$ limit, which corresponds to the ferromagnetic or antiferromagnetic phase depending on the sign of $J_1$. This model can be re-written in terms of parafermions after performing a non-local Fradkin-Kadanoff transformation.\cite{Fradkin1980, Fendley12} The parafermion with $k>2$ can be considered as a generalization of Majorana fermion for the transverse field Ising model ( $\mathbb{Z}_2$ clock model).\cite{Fradkin1980, Fendley12} For this model with open boundary condition, after performing a non-local Fradkin-Kadanoff transformation, there will be a parafermion zero mode left on the edge. As shown in Fig.\ref{defect_chain_dia} (c), in terms of twist defect chain model, this zero mode actually corresponds to the unpaired twist defect left on the boundary. Another limit is when $|J_1|<<|J_2|$, this is the disordered paramagnetic phase without any parafermion zero mode left on the boundary (Fig.\ref{defect_chain_dia} (d)). For the $\mathbb{Z}_k$ clock model, the Kramers-Wannier duality transformation in Eq.\eqref{KW_duality} exchanges the disordered paramagnetic phase and ordered phase. Under this duality transformation $D$,
\begin{equation}
DH(J_1,J_2)D^{-1}= H(J_2,J_1)
\end{equation}

The Hamiltonian Eq.\eqref{H_Wilson} has an ordered ferromagnetic / anti-ferromagnetic phase and a disordered paramagnetic phase and both of them are gapped phases. At the self-dual point, it turns out to be a gapless critical point protected by the additional duality symmetry.  The low energy excitation of this model is described by a conformal field theory (CFT) and  is closely related with the self-dual Sine-Gordon model,\cite{Lecheminant2002}
\begin{equation}
S=\int d^2\textbf{r}\frac{1}{2}(\partial_{\mu}\Phi)^2+g\cos(\sqrt{2\pi k}\Phi)+g\cos(\sqrt{2\pi k}\Theta)
\label{sdsg}
\end{equation}
where $\Phi$ is the bosonic field and $\Theta$ is the dual field. This model is invariant under the dual transformation $\Phi\leftrightarrow \Theta$ transformation and is always critical. In the renormalization group (RG) language, the two cosine terms are irrelevant when $k>4$ and therefore the self-dual Sine-Gordon model is the same as the Luttinger liquid at infrared (IR) limit with the central charge $c=1$. When $k=4$, the cosine terms are marginal, and the interaction will only change the compactification radius for the compact boson field. When $k<4$, the cosine terms are relevant, according to Zamolodchikov's c-theorem, the relevant perturbation will drive the model from the ultaviolet (UV) $c=1$ fixed point to the IR fixed point with $c<1$.\cite{Zamolodchikov1986}  When $k=2$, the self-dual Sine-Gordon model can be refermionized by introducing two Majorana fields. One of them will be gapped and the other one remains gapless with $c=1/2$. This is the Ising CFT and is the effective theory for critical $Z_2$ clock model (transverse field Ising model). When $k=3$, it corresponds to the three-state Potts CFT with $c=4/5$, and is identical to the critical $\mathbb{Z}_3$ clock model with ferromagnetic coupling.\cite{Ginsparg1988} The three-state Potts CFT can be considered as the deformation of $\mathbb{Z}_4$ parafermion CFT, which has $c=1$ and is the effective theory for the critical $\mathbb{Z}_3$ clock model with antiferromagnetic coupling.

\subsection{Wilson loop Hamiltonian with $2N-1$ twist defects}
\label{odd_num}
For the Wilson loop Hamiltonian with $2N-1$ number of twist defects defined in Eq.\eqref{Hamiltonian}, the Hamiltonian can still be written in terms of $\mathbb{Z}_k$ clock models with some twisted boundary term,
\begin{equation}
H = -J  \sum_{j=1}^{N-1} \left( \tau_j^{\dagger} \tau_{j+1} + \sigma_j + h.c. \right) - H_B
\label{odd_number_1}
\end{equation}
The boundary term $\mathcal{W}_{2N-1}$ [Fig.~\ref{defect_chain_dia}(b)] can be derived by computing the intersection with the neighboring Wilson loop and is proportional to $\sigma_N\tau_N^{\dag}\tau_1$. The coefficient in front of the boundary term depends on whether $k$ is even or odd and is fixed by the translational symmetry. The detail of the coefficient will be discussed in Sec. \ref{sec:trans}.

The Wilson loop Hamiltonian is translational invariant and thus the twisted $\mathbb{Z}_k$ clock model is still invariant under the Kramers-Wannier duality transformation. In this sense, this symmetry protects the criticality of the model. Since there are odd number of twist defects, the dimension of the total Hilbert space is equal to $k^{N-1/2}$ and the effective length for the twisted $\mathbb{Z}_k$ clock model is $L=N-1/2$. This will be useful in the numerical calculation later. The correspondence between the twist defect chain and the quantum $\mathbb{Z}_k$ clock model is summarized in Table.\ref{comparision}.

\begin{table}[htbp]
\centering
\begin{tabular}{ccc}
 $\mathbb{Z}_k$ clock model & Twofold defect chain \\\hline
Periodic boundary term & Even number of $\sigma_a$\\
 Duality twisted boundary term & Odd number of $\sigma_a$\\
 Duality symmetry (D)& $\mathcal{T}$\\
 Translation symmetry &  $\mathcal{T}^2$\\\hline
\end{tabular}
\caption{Comparision between the $\mathbb{Z}_k$ clock model and the twofold twist defect chain.}\label{comparision}
\end{table}

Actually, similar boundary condition has already been explored in the quantum $k$-state Potts model in Ref.\ \onlinecite{Schutz1993}.
The $k$-state Potts model with $k\leq 4$ can be constructed in terms of Temperley-Lieb Hamiltonian and can be exactly solved.
At the critical point (with $k\leq 4$), the model is invariant under the Kramers-Wannier duality transformation.
By constructing the duality symmetry operator in terms of Temperley-Lieb algebra generators,
Schultz noticed that there are two different kinds of duality symmetry operators which correspond to two classes of toroidal boundary terms \cite{Schutz1993}.
The first one is the traditional periodic boundary term in Eq.\eqref{even_chain} and the second one is duality twisted boundary term in Eq.\eqref{odd_number_1}.

In fact, in our twist defect chain, there is a simple way to interpret this duality symmetry operator in terms of twist defect,
which is equivalent to the translation symmetry operator $\mathcal{T}$ for the Wilson loop. As we mentioned in Eq.\eqref{translation},
it can be realized by a sequence of braiding operators. This is also illustrated in Fig.~\ref{b_even}, in the defect chain with $2N$ twist defects,
the translation operator involves $2N-1$ braiding moves, while for the $2N-1$ twist defects (shown in Fig.\ref{b_odd}),
$\mathcal{T}$ involves $2N-2$ braiding moves. In the next section, we will use the braiding operators to explicitly construct
the $\mathcal{T}$ operator/D operator.

\subsection{Translational symmetry $\mathcal{T}$}
\label{sec:trans}
As shown in Eq.\eqref{translation}, the $\mathcal{T}$/$D$ operator in the twist defect chain/$Z_k$ clock model can be realized by a product of braiding operators.
At critical point, the Hamiltonian is invariant under $\mathcal{T}$ operator.
In this section, we will use the braiding operator to construct $\mathcal{T}$ operator explicitly
for the twist defect chain and will also use the $\mathcal{T}$ operator to fix the coefficient of the boundary term.

The duality transformation operator for $k$-state Potts model (with two different boundary conditions)
has already been constructed using Temperley-Lieb algebra \cite{Schutz1993}.
Although the $k$-state Potts model and $\mathbb{Z}_k$ clock model are equivalent for $k=2,3$,
these two types of models are different for higher $k$ values. Therefore it is interesting to construct $\mathcal{T}$ operator systematically for
$\mathbb{Z}_k$ clock models and compare the result with that for $k$-state Potts model.

\subsubsection{Diagrammatical construction}

\begin{figure}[h]
	\centering
	\includegraphics[scale=.34]{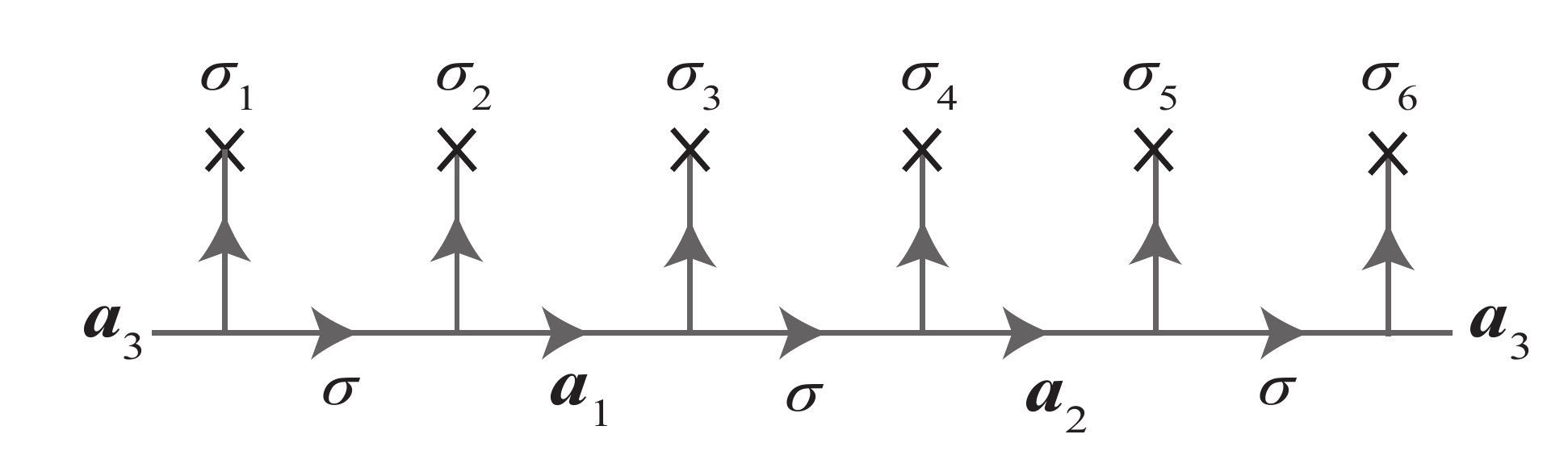}
	\caption{A quantum state in a system with 6 defects labeled by the internal fusion channels ${\bf a}_1$, ${\bf a}_2$, ${\bf a}_3$. The collection of quantum states $|{\bf a}_j\rangle$ forms an orthonormal basis of the Hilbert space.}
	\label{fusion_tree}
\end{figure}

\begin{figure}[h]
	\centering
	\includegraphics[scale=.34]{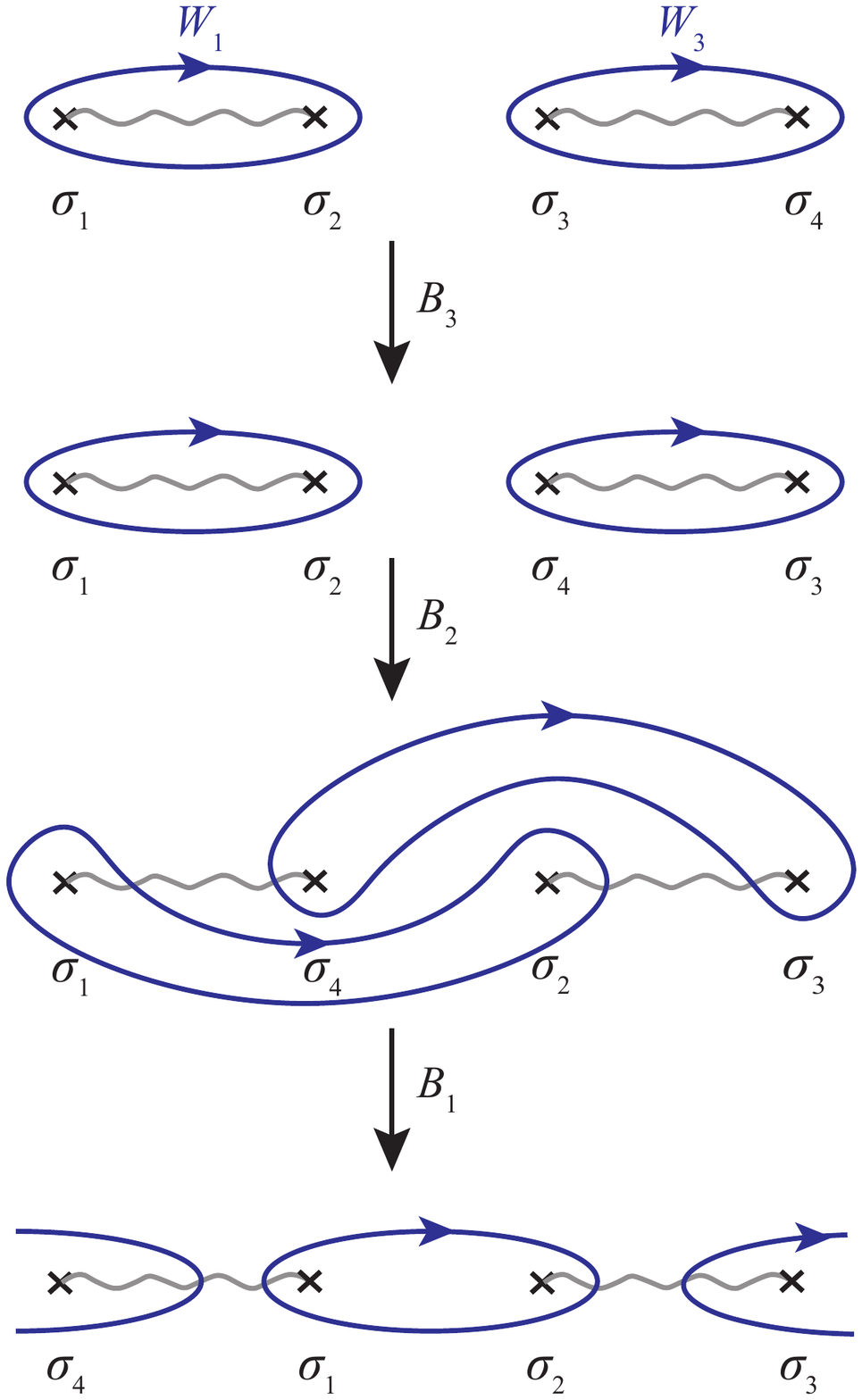}
	\caption{A schematic diagram for the translational operator $\mathcal{T}=B_{1}B_{2}B_{3}$ in a defect chain model with four twofold defects. After performing $\mathcal{T}$ operator on this chain, every defect moves one site to the right. The defect $\sigma_4$ on the right boundary moves to the left front. The Wilson loop operator is also shifted by one site.}
	\label{b_even}
\end{figure}

\begin{figure}[h]
	\centering
	\includegraphics[scale=.34]{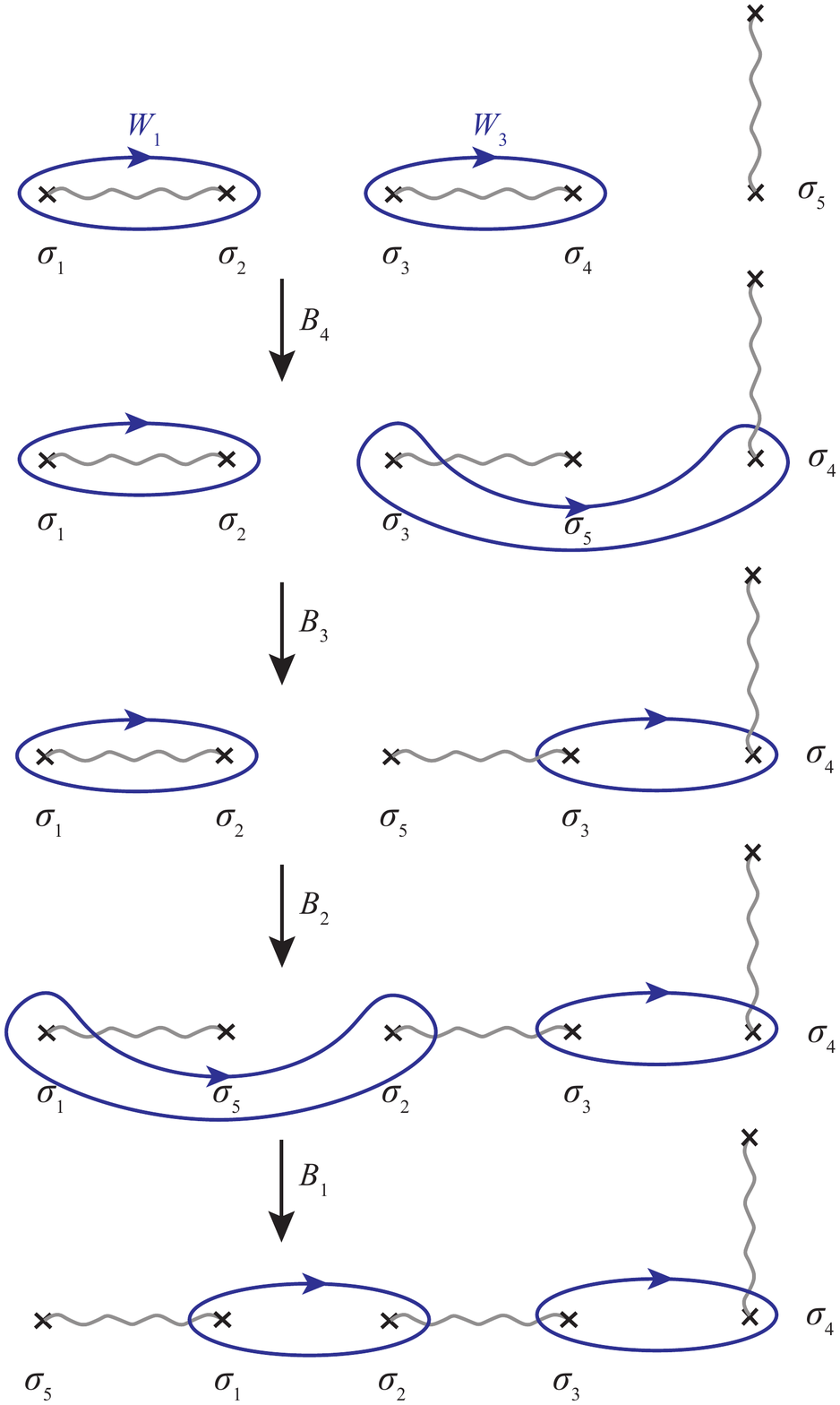}
	\caption{A schematic diagram for the translational operator $\mathcal{T}=B_{1}B_{2}B_{3}B_{4}$ in a defect chain model with five twofold defects. After performing $\mathcal{T}$ operator on this chain, every defect moves one site to the right. The defect $\sigma_5$ on the right boundary moves to the left front. The Wilson loop operator is also shifted by one site.}
	\label{b_odd}
\end{figure}

According to our results in previous paper,\cite{TeoRoyXiao13long} the braiding operator for twofold twist defect takes this form
\begin{eqnarray}
\nonumber &&B_{2j-1}^{(\textbf{a}_j, \textbf{a}_{j-1})}=e^{\frac{2\pi i}{k}(a_{j,2}-a_{j-1,2})[\frac{k}{2}-\frac{1}{2}(a_{j,2}-a_{j-1,2})]}\\
\nonumber &&\left[B_{2j}^{(\sigma_{j+1}, \sigma_{j})}\right]^{\textbf{a}_j}_{\textbf{a}^{\prime}_j}=\frac{e^{\frac{i\pi}{4}(k-1)}}{\sqrt{k}}e^{\frac{2\pi i}{k}(a^{\prime}_2-a_2)[\frac{k}{2}+\frac{1}{2}(a^{\prime}_2-a_2)]}\\
\label{braiding_eq}
\end{eqnarray}
where $B_{p}$ is the braiding operator for two neighboring twist defects at position $p$ and $p+1$.  For $B_{2j-1}$, it has the input and output channel fixed as abelian anyon $\textbf{a}_j$ and $\textbf{a}_{j+1}$ (Fig.~\ref{fusion_tree}), while $B_{2j}$ has the input and output channel as both twist defects. The superscript ${\bf a}_j$ and subscript ${\bf a}^{\prime}_j$ in Eq.\eqref{braiding_eq} are the internal channels and a braid operation can flip the anyon type of the internal channel. This braiding operator can be rewritten in the basis of $\mathbb{Z}_k$ rotors. When $k=2$, the braiding operator is equal to
\begin{eqnarray}
\nonumber &&B_{2j-1}=\frac{e^{\frac{\pi i}{4}}}{\sqrt{2}}(\sigma_j^z+e^{\frac{\pi i}{2}})\\
\nonumber &&B_{2j}=\frac{e^{-\frac{\pi i}{4}}}{\sqrt{2}}(\sigma_j^x\sigma_{j+1}^x+e^{\frac{\pi i}{2}})
\end{eqnarray}
 It is easy to check that
\begin{eqnarray}
\nonumber &&B_{2j}\sigma_j^zB_{2j}^{-1}=-i\sigma_j^x\sigma_j^z\sigma_{j+1}^x\\
&&B_{2j-1}(-i\sigma_j^x\sigma_j^z\sigma_{j+1}^x)B_{2j-1}^{-1}=\sigma_j^x\sigma_{j+1}^x
\end{eqnarray}
Combining these, we get the transformation $\mathcal{T}\sigma^z_j\mathcal{T}^{-1}=\sigma_j^x\sigma_{j+1}^x$. This can also be pictorially understood in Fig.~\ref{b_even} and Fig.~\ref{b_odd} which show $\mathcal{T}W_j\mathcal{T}^{-1}=W_{j+1}$. Similarly, this also shows that $\mathcal{T}\sigma_j^x\sigma_{j+1}^x\mathcal{T}^{-1}=\sigma_{j+1}^z$. This relation can be generalized to $k>2$ and we have
\begin{align}
&\mathcal{T}\mathcal{T}^{\dag}=\mathbb{I}\nonumber\\
&\mathcal{T}\sigma_j\mathcal{T}^{-1}=\tau_j\tau_{j+1}^{\dag},\quad (1\leq i\leq N-1)\nonumber\\
&\mathcal{T}\tau_j\tau_{j+1}^{\dag}\mathcal{T}^{-1}=\sigma_{j+1}, \quad (1\leq i\leq N-1)
\label{transl_symm}
\end{align}
The braiding operators for $k>2$ are listed in Appendix \ref{app:braiding}. Therefore $\mathcal{T}$ transformation shifts the Wilson loop Hamiltonian by one lattice spacing and corresponds to the duality transformation $D$ in $\mathbb{Z}_k$ clock model.


However, the above relation only holds for $\mathbb{Z}_k$ variables in the bulk and may not work for the boundary term, i.e.~the last term of Eq.\eqref{even_chain}. This is because on the boundary,
\begin{align}
& \mathcal{T} \sigma_{N} \mathcal{T}^{\dagger} =  Q \tau_N \tau_{1}^{\dagger}, \nonumber\\
& \mathcal{T} Q \tau_N \tau_{1}^{\dagger} \mathcal{T}^{\dagger} = \sigma_1.
\end{align}
There is an additional charge operator $Q$. Only when $Q$ takes a trivial value, i.e., $Q=1$, the above relation works for the boundary condition. For a general charge $Q$, when the $\mathcal{T}$ operator moves the twist defect along the chain, the charge $Q$ is also shifted with the twist defect, therefore, this model is not translational invariant under $\mathcal{T}$ operator.  One needs to carefully include boundary correction to $\mathcal{T}$, and define $\widetilde{\mathcal{T}}=\mathcal{T}X$ (to be discussed below), so that $\widetilde{\mathcal{T}}$ is an exact symmetry of the Hamiltonian.\cite{Schutz1993}

\subsubsection{Even number of twist defects}
For the case with even number of twist defects, one normally considers the periodic Hamiltonian as defined in Eq. \eqref{even_chain}, which amounts to choosing the boundary condition as $\tau_{N+1} = \tau_1$. More generally, one can consider boundary conditions with
\begin{equation}
	\tau_{N+1} = \omega^{-n} \tau_1, \ n=0,1,\cdots,k-1.
\end{equation}
which leads to a large set of Hamiltonians with different boundary conditions
\begin{equation}
\begin{split}
H^{(n)} =& -J \sum_{i=1}^{N-1} (\sigma_i + \tau_i \tau_{i+1}^{\dagger} + h.c.) \\
&  -J(\sigma_N + \omega^{n} \tau_N \tau_1^{\dagger} + h.c.),
\label{even_set}
\end{split}
\end{equation}
Define projection operator $P^{(n)}$ as
\begin{equation}
P^{(n)} = \frac{1}{k} (\sum_{i=0}^{k-1} \omega^{-ni} Q^i).
\end{equation}
It is easy to check that $P^{(n)}$ is a projection into the $Q=\omega^n$ sector. For instance, one can write $Q=\sum_{n=0}^{k-1}\omega^nP^{(n)}$. Mixed sector Hamiltonians are produced by mixing in one charge sector of every $H^{(n)}$, and defined as
\begin{equation}
\widetilde{H}^{(n)} = \sum_{m=0}^{k-1} P^{(m+n)}H^{(m)} P^{(m+n)}.
\end{equation}
So all charge sectors of the normal periodic Hamiltonian in Eq. \eqref{even_chain} are embedded into this set. Expanding the above equation, one gets
\begin{equation}
\begin{split}
\widetilde{H}^{(n)} =& -J \sum_{i=1}^{N-1} (\sigma_i + \tau_i \tau_{i+1}^{\dagger} + h.c.) \\
&  -J(\sigma_N + \omega^{-n} Q \tau_N \tau_1^{\dagger} + h.c.)
\end{split}
\end{equation}
If not for the additional phase factors of $\omega^{-n}$, the operator $\mathcal{T}$ would be an exact symmetry of $\widetilde{H}^{(n)}$.
To take into account that phase factor, following Ref.\ \onlinecite{Schutz1993}, one could choose a boundary correction term $X=\tau_N^nV$,
where the global operator $V$ is defined through its action on the $\mathbb{Z}_k$ variables,
\begin{align}
\label{V_const}
	V^2 & =\mathbb{I} \nonumber \\
	V\sigma_i V &= \sigma_i^{\dagger} \nonumber \\
	V\tau_i V &= \tau_i^{\dagger}
\end{align}
Actually, there is an explicit matrix representation for $V$. First, we consider $V_i$ which only acts on the $\sigma_i$, $\tau_i$ sector. Let $V_i=(v_{ab})$ where $a,b=0,1,...,k-1$. Take $v_{ab}=1$ for $a+b=0$ mod $k$ and $0$ otherwise. In other words, $V_i$ is the matrix
\begin{align}
\begin{pmatrix}
1 & 0 & 0 & \cdots & 0 & 0\\
0 & 0 & 0 & \cdots & 0 & 1\\
0 & 0 & 0 & \cdots & 1 & 0\\
\cdot & \cdot & \cdot & \cdots & \cdot & \cdot\\
0 & 0 & 1 & \cdots & 0 & 0\\
0 & 1 & 0 & \cdots & 0 & 0
\end{pmatrix}
\end{align}
Finally we can take the tensor product $V=V_1\otimes ... \otimes V_N$ which satisfies Eq.\eqref{V_const}.

For $\widetilde{\mathcal{T}}=\mathcal{T}\tau_N^nV$, now it is easy to check that
\begin{align}
& \widetilde{\mathcal{T}} \sigma_i^{\dagger} \widetilde{\mathcal{T}}^{-1} = \tau_i \tau_{i+1}^{\dagger} \quad (1 \le i \le N-1),\\
& \widetilde{\mathcal{T}} \tau_i \tau_{i+1}^{\dagger} \widetilde{\mathcal{T}}^{-1} = \sigma_{i+1}^{\dagger} \quad (1 \le i \le N-1),\\
& \widetilde{\mathcal{T}} \sigma_{N}^{\dagger} \widetilde{\mathcal{T}}^{-1} =  \omega^{-n} Q \tau_N \tau_{1}^{\dagger}, \\
& \widetilde{\mathcal{T}} \omega^{-n} Q \tau_N \tau_{1}^{\dagger} \widetilde{\mathcal{T}}^{-1} = \sigma_1^{\dagger}.
\end{align}
Therefore, we have $\widetilde{\mathcal{T}} \widetilde{H}^{(n)} \widetilde{\mathcal{T}}^{-1} = \widetilde{H}^{(n)}$, which means $\widetilde{\mathcal{T}}$ is an exact duality symmetry of mixed sector Hamiltonian $\widetilde{H}^{(n)}$, and $\widetilde{\mathcal{T}}^2$ is a translational operator in the $\mathbb{Z}_k$ clock model.

\subsubsection{Odd Number of twist defects}
The construction of $\widetilde{\mathcal{T}}$ is very similar to the previous case. Define $\mathcal{T} = B_1 B_2 \cdots B_{2N-2}$, and $\widetilde{\mathcal{T}}=\mathcal{T}X$, where $X$ is a boundary correction to be clarified.
Using the braiding operators $B$ discussed in Appendix. \ref{app:braiding}, one can show that
\begin{align}
& \mathcal{T} \sigma_i \mathcal{T}^{\dagger} = \tau_i \tau^{\dagger}_{i+1} \quad (1 \le i \le N-1),\\
& \mathcal{T} \tau_i \tau^{\dagger}_{i+1} \mathcal{T}^{\dagger} = \sigma_{i+1} \quad (1 \le i \le N-2),\\
& \mathcal{T} \tau_{N-1} \tau_{N}^{\dagger} \mathcal{T}^{\dagger} =  e^{i\phi_k} Q^{\dagger} \tau_N^{\dagger} \tau_{1} \sigma_{N},\label{DT_phase}\\
& \mathcal{T} e^{i\phi_k} Q^{\dagger} \tau_N^{\dagger} \tau_{1} \sigma_{N} \mathcal{T}^{\dagger} = \sigma_1^{\dagger}.
\end{align}
where $\phi_k$ is a phase determined by Eq. \eqref{DT_phase} and is equal to
$\phi_3=2\pi/3$, $\phi_4=-\pi/4$, $\phi_5=4\pi/5$, and $\phi_6=-\pi/6$. Once again, the additional terms in the last two lines need to be matched with a properly chosen $H_B$, so that $\widetilde{\mathcal{T}}$ is an exact symmetry of the Hamiltonian.

In order to construct the mixed sector Hamiltonians that commute with $\widetilde{\mathcal{T}}$, we consider the following duality twisted Hamiltonians with different boundary conditions
\begin{equation}
\begin{split}
H^{(m)} =& -J \sum_{i=1}^{N-1} (\sigma_i + \tau_i \tau_{i+1}^{\dagger} + h.c.) \\
&  -J(\omega^{m} e^{-i\phi_k} \sigma_N^{\dagger} \tau_N \tau_1^{\dagger} + h.c.).\label{DT_Hamiltonians}
\end{split}
\end{equation}
where $m=0,1,\cdots,k-1$. Unlike the even number case, here the Hamiltonians $H^{(m)}$ are not actually independent, as they are related by a local unitary transformation
\begin{equation}
	H^{(m+1)} = \tau_N H^{(m)} \tau^{\dagger}_N, \quad  Q = \omega \tau_N Q \tau^{\dagger}_N.  \label{Unitary_relation}
\end{equation}
To elaborate on its meaning, let's assume that we have an eigenstate $|\alpha, q\rangle$ of $H^{(m)}$ and $Q$, which satisfies $H^{(m)} |\alpha, q\rangle = E_{\alpha} |\alpha, q\rangle$ and $Q |\alpha, q\rangle = q |\alpha, q\rangle$. Then Eq. \eqref{Unitary_relation} says that $\tau_N |\alpha, q\rangle$ is an eigenstate of $H^{(m+1)}$ and $Q$, since $H^{(m+1)} \tau_N |\alpha, q\rangle = E_{\alpha} \tau_N |\alpha, q\rangle$ and $Q \tau_N |\alpha, q\rangle = \omega q \tau_N |\alpha, q\rangle$.
This essentially means Hamiltonians $H^{(m)}$ of different $m$ are all equivalent, up to some changes to the charge sector labels.

We define the mixed sector Hamiltonians as
\begin{equation}
\widetilde{H}^{(m)}= \sum_{l=0}^{k-1} P^{(l+m)}H^{(l)} P^{(l+m)}.
\end{equation}
which expands to
\begin{equation}
\label{H_odd_mixed}
\begin{split}
\widetilde{H}^{(m)}=& -J \sum_{i=1}^{N-1} (\sigma_i + \tau_i \tau_{i+1}^{\dagger} + h.c.) \\
&  -J(\omega^{-m} e^{-i\phi_k} Q \sigma_N^{\dagger} \tau_N \tau_1^{\dagger} +h.c.)
\end{split}
\end{equation}

Similarly, one can show that for $\widetilde{\mathcal{T}}=\mathcal{T}(\sigma_N^{\dagger})^mV$,
\begin{align}
& \widetilde{\mathcal{T}} \sigma_i^{\dagger} \widetilde{\mathcal{T}}^{-1} = \tau_i \tau_{i+1}^{\dagger} \quad (1 \le i \le N-1),\\
& \widetilde{\mathcal{T}} \tau_i \tau_{i+1}^{\dagger} \widetilde{\mathcal{T}}^{-1} = \sigma_{i+1}^{\dagger} \quad (1 \le i \le N-2),\\
& \widetilde{\mathcal{T}} \tau_{N-1} \tau_{N}^{\dagger} \widetilde{\mathcal{T}}^{-1} = \omega^{-m} e^{-i\phi_k} Q \sigma_{N}^{\dagger} \tau_N \tau_{1}^{\dagger},\\
& \widetilde{\mathcal{T}} \omega^{-m} e^{-i\phi_k} Q \sigma_{N}^{\dagger} \tau_N \tau_{1}^{\dagger} \widetilde{\mathcal{T}}^{-1} = \sigma_1^{\dagger}.
\end{align}
Therefore, we have $\widetilde{\mathcal{T}} \widetilde{H}^{(m)} \widetilde{\mathcal{T}}^{-1} = \widetilde{H}^{(m)}$, which means $\widetilde{\mathcal{T}}$ is an exact duality symmetry of $\widetilde{H}^{(m)}$. $\widetilde{\mathcal{T}}^2$ can also be interpreted as a special ``translational operator", in the sense that it correctly translates a local term which is far from the bounday, and applying it $(N-1/2)$ number of times on a local term will return it back to itself, where the $1/2$ factor comes from the twisted boundary condition. Notice that $\widetilde{\mathcal{T}}$ does not commute with $Q$.

Numerically, we are interested in the CFT content of $H^{(0)}$, but since $H^{(0)}$ does not commute with $\widetilde{\mathcal{T}}$, it is more favorable to work with $\widetilde{H}^{(m)}$, from which we can extract the ``momentum" eigenvalues of $\widetilde{\mathcal{T}}^2$ as well. Due to Eq. \eqref{Unitary_relation}, for $\widetilde{H}^{(m)}$ with a fixed $m$, all these $k$ different charge sectors (labelled by charge operator $Q$ ) have the same energy spectrum. However, $\widetilde{H}^{(m)}$ with different $m$ can have different energy spectrum and corresponds to different charge sector in $H^{(0)}$. Therefore, to obtain the low lying energy levels of all the charge sectors of $H^{(0)}$, we need to solve for the low lying energy levels of $k$ different $\widetilde{H}^{(m)}$. To distinguish the results from the even-defect chains, when summarizing the conformal dimensions into tables, we describe the charge sectors using label $m$, in correspondence to the fact that we are using mixed sector Hamiltonians $\widetilde{H}^{(m)}$.

\section{Numerical methods and numerical results}
\label{numerics}
To determine the properties of the CFTs underlying the Wilson loop Hamiltonian in Eq.(2.2),
we extract the conformal dimensions $h$ and $\bar{h}$ using finite and infinite density matrix renormalization group
(DMRG/iDMRG) \cite{White1992}, and exact diagonalization (ED) methods.
The DMRG calculations are based on the open-source C++ library ITensor \cite{ITensor}.

\subsubsection{Wilson loop Hamiltonian with $2N$ twist defects}
Twist defect chains with $2N$ twist defects in Eq. \eqref{even_chain} correspond to critical $\mathbb{Z}_k$ clock models with length $L=N$,
and have difference CFTs at criticality. In order to uncover the contents of the CFTs, except for a few cases which allow analytical solutions,
one needs to perform numerical calculations on the energy spectrum of the critical system as explained below.

It has been shown that the energy spectrum of a critical chain with finite length $L$ with periodic boundary conditions
obeys \cite{Affleck1986, Blote1986, CARDY1986}
\begin{eqnarray}\label{E_spectra}
\nonumber E&=&\epsilon_{\infty}L - \frac{\pi v c}{6L} + \frac{2\pi v}{L}(h+\bar{h}+n+\bar{n}) + O(L^{-2})\\
&=& E_0+ \frac{2\pi v}{L}(h+\bar{h}+n+\bar{n}) + O(L^{-2})
\label{cft_spec}
\end{eqnarray}
where $\epsilon_{\infty}$ is the energy density of the ground state in the limit of $L\to\infty$;
$v$ is the sound velocity; $c$ is the central charge. These three parameters
can be pinned down by using DMRG method with high accuracy. The results are listed in Table~\ref{ZkData} and the detail for the numerical calculation is explained in Appendix~\ref{gs_energy}. $E_0=\epsilon_{\infty}L - \pi v c/6L$, up to small correction in order of $O(L^{-2})$, is the ground state energy; $h$ is the holomorphic conformal dimension of the primary field and
$\bar{h}$ is the anti-holomorphic counterpart; and $n$ and $\bar{n}$ are non-negative integers marking the energy levels.
Besides, the momentum quantum numbers are related to the conformal dimensions of the primary (and descendant) fields as
\begin{equation}
P = \frac{2\pi}{L}(h+n-\bar{h}-\bar{n}).
\label{even_rescaled_P}
\end{equation}
As a side note, in ED calculations, $P$ can be trivially obtained through the eigenvalues of the translation operator.
One usually shifts the eigenstates by one site, and the resulting phase factors would lead to the quantized momenta.
The only complication arises because of the even and odd pattern of the critical anti-ferromagnetic $Z_3$ and $Z_5$ clock models.\cite{Li2015}
For these cases with different system sizes, the eigenstates' momenta are only consistent
if one calculates the phase factors through a translation by two sites, while translation by one site does not produce meaningful results.

\begin{table}[htbp]
\centering
\begin{tabular}{c|c|c|c|c}
$k$ & coupling & $c$ & $\epsilon_{\infty}$ & $v$ \\\hline
$3$ & F & $0.8$ & $-2.43599110$ & $2.59802$ \\
$3$ & AF & $1$ & $-1.81607175$ & $1.29901$ \\
\hline
$4$ & F/AF & $1$ & $-2.54647904$ & $1.99992$ \\
\hline
$5$ & F & $1$ & $-2.7184737$ & $1.6811$ \\
$5$ & AF & $1$ & $-2.68272$ & $1.5206$ \\
\hline
$6$ & F/AF & $1$ & $-2.880358$ & $1.471$ \\
\end{tabular}
\caption{Central charge, ground states' energy per site, and sound velocity of $\mathbb{Z}_3$, $\mathbb{Z}_4$, $\mathbb{Z}_5$, and $\mathbb{Z}_6$ clock models.}
\label{ZkData}
\end{table}

Once we have all the necessary parameters, we can calculate the conformal dimensions based on the rescaled energy
\begin{equation}
E_R \equiv (h+\bar{h}+n+\bar{n}) = \frac{L}{2\pi v} (E - E_0) + O(L^{-1}),
\label{even_rescaled_E}
\end{equation}
and the rescaled momenta $\frac{L}{2\pi} P$, since
\begin{align}
& h+n = \frac{1}{2}(E_R + \frac{L}{2\pi} P) + O(L^{-1}),\nonumber\\
& \bar{h}+\bar{n} = \frac{1}{2}(E_R - \frac{L}{2\pi} P) + O(L^{-1}).
\label{even_rescaled_E}
\end{align}
Because of the finite size correction at the order of $O(L^{-1})$, polynomial extrapolation in terms of $1/L \to 0$ is often needed for small system sizes. The detail for this calculation is shown in Appendix \ref{rescaled_energy}.

In the Table \ref{Scaled_K3_F_even},\ref{Scaled_K3_AF_even},\ref{Scaled_K4_even},\ref{Scaled_K5_even},\ref{Scaled_K6_even},
we show the results for the $k=3,4,5,6$ critical chain with even number of twist defects with both ferromagnetic/antiferromagnetic coupling.
$h$ and $\bar{h}$ extracted from energy spectrum match up with the known CFT results for critical $\mathbb{Z}_k$ clock models and are summarized in Table~\ref{summary_even_odd}.
In the following section, we will study the energy spectrum for the odd number of twist defects and compare the results with the even case.

\begin{table}[htbp]
\centering
\begin{tabular}{c|c|c|c}
$q$ & $E_R$ & $\frac{L}{2\pi} P$ & $(h+n, \bar{h}+\bar{n})$ \\\hline
$0$ & 0 & 0 & $(0,0)$ \\
$0$ & 0.80304 & 0 & $(\frac{2}{5},\frac{2}{5})$ \\
$0$ & 1.79620 & $\pm 1$ & $(\frac{2}{5},\frac{7}{5})$, $(\frac{7}{5},\frac{2}{5})$ \\
$0$ & 1.80524 & $\pm 1$ & $(\frac{2}{5},\frac{7}{5})$, $(\frac{7}{5},\frac{2}{5})$ \\
$0$ & 1.99987 & $\pm 2$ & $(2,0)$, $(0,2)$ \\
\hline
$1,2$ & 0.13341 & 0 & $(\frac{1}{15},\frac{1}{15})$ \\
$1,2$ & 1.13378 & $\pm 1$ & $(\frac{1}{15},\frac{16}{15})$, $(\frac{16}{15},\frac{1}{15})$\\
$1,2$ & 1.33391 & 0 & $(\frac{2}{3},\frac{2}{3})$ \\
$1,2$ & 2.13351 & $\pm 2$ & $(\frac{1}{15},\frac{31}{15})$, $(\frac{31}{15},\frac{1}{15})$
\end{tabular}
\caption{Conformal dimensions of primary and descendant fields of the ferromagnetic $\mathbb{Z}_3$ clock model with even number of twist defects.
$q$ above is a label for the charge sector, with $e^{2 q \pi i/k}$ as the eigenvalue of $Q$.
The results shown above are obtained using ED for $N=5,6,7,8,9,10,11,12,13,14$.
Rescaled energies higher than 2.13351 are not well resolved for the system sizes available.}
\label{Scaled_K3_F_even}
\end{table}

\begin{table}[htbp]
\centering
\begin{tabular}{c|c|c|c}
$q$ & $E_R$ & $\frac{L}{2\pi} P$ & $(h+n, \bar{h}+\bar{n})$ \\\hline
$0$ & 0 & 0 & $(0,0)$ \\
$0$ & 1.00027 & $\pm 1$ & $(1,0)$, $(0,1)$  \\
$0$ & 1.49975 & 0 & $(\frac{3}{4},\frac{3}{4})$ \\
$0$ & 1.50268 & 0 & $(\frac{3}{4},\frac{3}{4})$ \\
\hline
$1,2$ & 0.16667 & 0 & $(\frac{1}{12},\frac{1}{12})$ \\
$1,2$ & 0.66675 & 0 & $(\frac{1}{3},\frac{1}{3})$\\
$1,2$ & 1.16728 & $\pm 1$ & $(\frac{1}{12},\frac{13}{12})$, $(\frac{13}{12},\frac{1}{12})$ \\
$1,2$ & 1.67321 & $\pm 1$ & $(\frac{1}{3},\frac{4}{3})$, $(\frac{4}{3},\frac{1}{3})$
\end{tabular}
\caption{Conformal dimensions of primary and descendant fields of the anti-ferromagnetic $\mathbb{Z}_3$ clock model with even number of twist defects.
$q$ above is a label for the charge sector, with $e^{2 q \pi i/k}$ as the eigenvalue of $Q$. Here we only show the lowest 10 excitations.
The results shown above are obtained using ED for $N=6,8,10,12,14$.}
\label{Scaled_K3_AF_even}
\end{table}

\begin{table}[htbp]
\centering
\begin{tabular}{c|c|c|c}
$q$ & $E_R$ & $\frac{L}{2\pi} P$ & $(h+n, \bar{h}+\bar{n})$ \\\hline
$0$ & 0 & 0 & $(0,0)$ \\
$0$ & 0.99999 & 0 & $(\frac{1}{2},\frac{1}{2})$ \\
$0$ & 1.24986 & $\pm 1$ & $(\frac{1}{8},\frac{9}{8})$, $(\frac{9}{8},\frac{1}{8})$ \\
$0$ & 1.99896 & $\pm 1$, $\pm 2$ & $(\frac{3}{2},\frac{1}{2})$, $(\frac{1}{2},\frac{3}{2})$, $(2,0)$, $(0,2)$\\
\hline
$1,3$ & 0.12500 & 0 & $(\frac{1}{16},\frac{1}{16})$ \\
$1,3$ & 1.12485 & 0,$\pm 1$ & $(\frac{9}{16},\frac{9}{16})$, $(\frac{1}{16},\frac{17}{16})$, $(\frac{17}{16},\frac{1}{16})$ \\
\hline
$2$ & 0.25000 & 0 & $(\frac{1}{8},\frac{1}{8})$ \\
$2$ & 0.99999 & 0 & $(\frac{1}{2},\frac{1}{2})$ \\
$2$ & 1.24986 & $\pm 1$ & $(\frac{1}{8},\frac{9}{8})$, $(\frac{9}{8},\frac{1}{8})$ \\
$2$ & 1.99896 & $\pm 1$, $\pm 2$ & $(\frac{3}{2},\frac{1}{2})$, $(\frac{1}{2},\frac{3}{2})$, $(2,0)$, $(0,2)$
\end{tabular}
\caption{Conformal dimensions of primary and descendant fields of the ferromagnetic/antiferromagnetic $\mathbb{Z}_4$ clock model with even number of twist defects.
$q$ above is a label for the charge sector, with $e^{2 q \pi i/k}$ as the eigenvalue of $Q$.
The results shown above are obtained using ED for $N=5,6,7,8,9,10,11,12$.
}\label{Scaled_K4_even}
\end{table}

\begin{table}[htbp]
\centering
\begin{tabular}{c|c|c|c|c}
$q$ & $E^F_R$ & $E^F_R$ & $\frac{L}{2\pi} P^{F/AF}$ & $(h+n, \bar{h}+\bar{n})^{F/AF}$ \\\hline
$0$ & 0 & 0 & 0 & $(0,0)$ \\
$0$ & 1.0014  & 1.0004 & $\pm 1$ & $(1,0)$, $(0,1)$ \\
\hline
$1,4$ & 0.0998 & 0.1001 & 0 & $(\frac{1}{20},\frac{1}{20})$ \\
$1,4$ & 1.0959 & 1.1050 & $\pm 1$ & $(\frac{1}{20},\frac{21}{20})$, $(\frac{21}{20},\frac{1}{20})$\\
\hline
$2,3$ & 0.3939 & 0.4010 & 0 & $(\frac{1}{5},\frac{1}{5})$ \\
$2,3$ & 0.8930 & 0.8979 & 0 & $(\frac{9}{20},\frac{9}{20})$
\end{tabular}
\caption{Conformal dimensions of the primary and descendant fields of ferromagnetic (F) and anti-ferromagnetic (AF) $\mathbb{Z}_5$ clock models.
$q$ above is a label for the charge sector, with $e^{2 q \pi i/k}$ as the eigenvalue of $Q$.
The ferromagnetic results shown above are obtained using ED for $N=5,6,7,8,9,10,11,12$, while the anti-ferromagnetic results are obtained using ED for $N=4,6,8,10,12$.
Notice that the conformal dimensions for F and AF cases are identical.
}\label{Scaled_K5_even}
\end{table}

\begin{table}[htbp]
\centering
\begin{tabular}{c|c|c|c}
$q$ & $E_R$ & $\frac{L}{2\pi} P$ & $(h+n, \bar{h}+\bar{n})$ \\\hline
$0$ & 0 & 0 & $(0,0)$ \\
$0$ & 0.9997 & $\pm 1$ & $(1,0)$, $(0,1)$ \\
\hline
$1,5$ & 0.0834 & 0 & $(\frac{1}{24},\frac{1}{24})$ \\
\hline
$2,4$ & 0.3342 & 0 & $(\frac{1}{6},\frac{1}{6})$ \\
\hline
$3$ & 0.7561 & 0 & $(\frac{3}{8},\frac{3}{8})$
\end{tabular}
\caption{Conformal dimensions of primary and descendant fields of the ferromagnetic/antiferromagnetic $\mathbb{Z}_6$ clock model with even number of twist defects.
$q$ above is a label for the charge sector, with $e^{2 q \pi i/k}$ as the eigenvalue of $Q$.
The results shown above are obtained from ED results of $N=4,5,6,7,8,9,10$.
}\label{Scaled_K6_even}
\end{table}


\subsubsection{Wilson loop Hamiltonian with $2N-1$ twist defects}
The Wilson loop Hamiltonian with $2N-1$ twist defects corresponds to the twisted $\mathbb{Z}_k$ clock model in Eq.\eqref{H_odd_mixed}.
Although the twisted $\mathbb{Z}_k$ clock model still has $N$ sites, the effective length is $L=N-1/2$ and the energy spectra of the twisted
$\mathbb{Z}_k$ clock model is  described by Eq.\eqref{cft_spec},
where the parameters $\epsilon_{\infty}$, $c$ and $v$ are the same as in the even number case.

As with the previous case, $(h+\bar{h}+n+\bar{n})$ can be obtained by calculating the rescaled energy $E_R$, which is now defined in this way,
\begin{align}
E_R &\equiv (h+\bar{h}+n+\bar{n})\nonumber\\
& = \frac{ (N-\frac{1}{2})^2( \frac{E}{N-\frac{1}{2}} - \epsilon_{\infty})+\frac{\pi v c}{6} } {2\pi v} + O(N^{-1})
\label{odd_rescaled_E}
\end{align}
However complication arises for the calculation of $(h+n-\bar{h}-\bar{n})$. First of all, under the duality twist boundary conditions,
it is not known a priori whether there exists a
relation between the momenta and $(h+n-\bar{h}-\bar{n})$ as in Eq. \eqref{even_rescaled_P}.
Secondly, assuming the the same relations holds, there no longer exists a simple translation operator,
where one can shift the eigenstates by one or two sites in ED, and find the momenta through the phase factors.
The system as defined in Eq. \eqref{odd_number_1} has a translational operation given by $\widetilde{\mathcal{T}}$,
which is built by consecutive multiplication of $B$-operators and commutes with the Hamiltonian. We will calculate the eigenvalues of $\widetilde{\mathcal{T}}$ and extract the  ``momenta" of a system of length $N-1/2$.
However, since there is an overall phase ambiguity in the definition of $\widetilde{\mathcal{T}}$, the ``momentum" quantum numbers are quantized up to an unknown additive constant which changes in each charge sector and also depends on each system size.
Therefore we can only fix the difference of the
``momentum" quantum numbers of $\widetilde{\mathcal{T}}$ (denoted as $\Delta P$) between any excited state and the lowest energy eigenstate in identical
charge sector and of the same system size. Due to this overall phase ambiguity, it is not possible to pin down a unique combination of $(h+n,\bar{h}+\bar{n})$.

However, in most of the cases, it turns out the decomposition into $(h+n,\bar{h}+\bar{n})$ is quite simple. We will show that when $k\neq 4$, in each charge sectors, only the holomorphic part ($h$) or the anti-holomorphic part ($\bar{h}$) has a twist with the rest part remains the same. This result matches up with the $\mathbb{Z}_2$ orbifold CFT. The $k=4$ case is more complicated since we don't find any known orbifold CFT which precisely has the same excitation spectrum. Nevertheless, we still manage to show that these new excitations in $k=4$ case should be related with $\mathbb{Z}_4$ twist fields.


\subsection{$k=3$, Ferromagnetic}
The $\mathbb{Z}_3$ clock model with ferromagnetic coupling at critical point can be described by the three-state Potts CFT. It has a block-diagonal modular invariant partition function, \cite{Gehlen1986,CARDY1986b}
\begin{align}
Z=|\chi_0+\chi_3|^2+|\chi_{\frac{2}{5}}+\chi_{\frac{7}{5}}|^2+2|\chi_{\frac{1}{15}}|^2+2|\chi_{\frac{2}{3}}|^2
\end{align}
where $\chi_h$ denotes the character for each primary field with conformal dimension $h$. As shown in Table \ref{Scaled_K3_F_even}, $h$ obtained from the energy spectrum difference is consistent with the CFT prediction.

Once we consider odd-defect chain at critical point, there will be some new excitations in the low energy spectrum. In Table \ref{Scaled_K3_F_odd}, we present the rescaled energy spectrum and momentum difference. From these numerical data, we can calculate the possible combination $(h+n, \bar{h}+\bar{n})$ and we find two new excitations with conformal dimensions equal to $1/40$ and $1/8$. They are not in the original three-state Potts CFT but can be found in $\mathcal{M}(5,6)$ minimal model.

Actually, the three-state Potts CFT can be considered as a subset of $\mathcal{M}(5,6)$ minimal model (tetracritical Ising CFT), which includes all the ten primary fields and has a diagonal modular invariant partition function $Z=\sum_{i=1}^{10}|\chi_i|^2$. These ten primary fields have conformal dimension $h=0, 1/8, 2/3, 13/8, 2/5, 1/40, 1/15, 21/40, 7/5, 3$.\cite{Gehlen1986,CARDY1986b,bigyellowbook} The two CFTs are connected through $\mathbb{Z}_2$ orbifolding and the $\mathcal{M}(5,6)$ minimal model involves some new twist field operators. There is a simple way to understand this $\mathbb{Z}_2$ orbifold:\cite{Ginsparg1988} the three-state Potts CFT ($\mathbb{Z}_3$ parafermion CFT) is defined by the coset,
\begin{align}
\frac{SU(2)_3}{U(1)_3}=(G_2)_1\times\overline{SU(3)}_1=\langle 1,\tau\rangle\times \langle 1,\overline{s},\overline{s^2}\rangle
\end{align}
where $(G_2)_1$ refers to the exceptional Lie group $G_2$ at level $1$ and it contains $1$ and the Fibonacci anyon $\tau$ with conformal dimension $h_{\tau}=2/5$. Here $\overline{SU(3)_1}$ means the time reversal or anti-holomorphic part of $SU(3)$ with the reverse propagating direction.  Notice that $SU(3)_1$ CFT contains three primary fields $1$, $s$ and $s^2$ with $h_{s,s^2}=1/3$.  Therefore, the three-state Potts CFT can be understood as the tensor product between $(G_2)_1$ and $\overline{SU(3)_1}$ with $2\times 3=6$ primary fields and has central charge $c=14/5-2=4/5$.

The abelian $SU(3)_1$ CFT has $\mathbb{Z}_2$ symmetry. After orbifolding this $\mathbb{Z}_2$ symmetry, it becomes $SU(2)_4$ CFT which has five primary fields with conformal dimension $h=0,~ 1/8, ~ 1/3,~ 5/8,~ 1$.\cite{BarkeshliBondersonChengWang14,Teo2015,ChenRoyTeoRyu17} Among them, there are two $\mathbb{Z}_2$ twist fields with $h=1/8,~5/8$. Combined with $(G_2)_1$ sector, this new CFT has ten primary fields and has similar structure as the $\mathcal{M}(5,6)$ minimal model.

Coming back to Table~\ref{Scaled_K3_F_odd}, we observe that $\bar{h}$ is still the same as the original three-state Potts CFT, while $h$ is new and comes from the $\mathbb{Z}_2$ twist fields in $\mathcal{M}(5,6)$ minimal model. We will show that similar behavior  occurs for other cases except $k=4$ model.

\begin{table}[htbp]
\centering
\begin{tabular}{c|c|c|c}
$m$ & $E_R$ & $ \frac{L-1/2}{2\pi}\Delta P$ & $(h+n, \bar{h}+\bar{n})$ \\\hline
$0$ & 0.12499 & --- --- & $(\frac{1}{8},0)$\\
$0$ & 0.42533 & $-\frac{1}{2}$ & $(\frac{1}{40},\frac{2}{5})$ \\
$0$ & 0.92317 & $0$ & $(\frac{21}{40},\frac{2}{5})$ \\
\hline
$1,2$ & 0.09161 & --- --- & $(\frac{1}{40},\frac{1}{15})$\\
$1,2$ & 0.59202 & $\frac{1}{2}$ & $(\frac{21}{40},\frac{1}{15})$ \\
$1,2$ & 0.79175 & $-\frac{1}{2}$ & $(\frac{1}{8},\frac{2}{3})$
\end{tabular}
\caption{Conformal dimensions of primary fields of the ferromagnetic $\mathbb{Z}_3$ clock model with odd number of twist defects.
$m$ above is a label for the mixed sector Hamiltonian $\widetilde{H}^{(m)}$, or the $Q=e^{2 m \pi i/k}$ charge sector for $H^{(0)}$.
$\Delta P$ is the difference of the ``momentum" quantum numbers between any excited eigenstate and
the lowest energy eigenstate in the same charge sector and of the same system size.
The results shown above are obtained using ED for $N=7,8,9,10,11,12,13,14$.}
\label{Scaled_K3_F_odd}
\end{table}

\subsection{$k=3$, Anti-Ferromagnetic \& $k\geq 5$}

For the even-defect chain with $k=3$, i.e., $\mathbb{Z}_3$ clock model, if the coupling is antiferromagnetic, the critical point is described by $U(1)_3$ CFT with $\mathbb{Z}_2$ charge-conjugation symmetry. The conformal dimension for this CFT is equal to $r^2/12$  with $0\leq r<6$ and $r\in\mathbb{Z}$. In Table \ref{Scaled_K3_AF_even}, we present the numerical results for  $h$ and $\bar{h}$. For each excited state, $h$ and $\bar{h}$ are always the same, suggesting that the partition function takes a diagonal form.

For the odd-defect chain shown in Table \ref{Scaled_K3_AF_odd}, we observe that the ground state has energy shifted by $1/16$, which is the same as the conformal dimension for $\mathbb{Z}_2$ twist field in $U(1)_3/\mathbb{Z}_2$ CFT (The detail for $\mathbb{Z}_2$ orbifold CFT is shown in Appendix \ref{U_1_orb}).\cite{bigyellowbook,Ginsparg1988} ). Moreover, $h$ and $\bar{h}$ do not come in pairs. $h$ is still the same as the original $U(1)_3$ CFT, while $\bar{h}$ is coming from the $\mathbb{Z}_2$ twist field.


\begin{table}[htbp]
\centering
\begin{tabular}{c|c|c|c}
$m$ & $E_R$ & $\frac{L-1/2}{2\pi}\Delta P$ & $(h+n, \bar{h}+\bar{n})$ \\\hline
$0$ & 0.06250 & --- --- & $(0,\frac{1}{16})$\\
$0$ & 0.56249 & $-\frac{1}{2}$ & $(0,\frac{9}{16})$\\
$0$ & 0.81244 & $\frac{3}{4}$ & $(\frac{3}{4},\frac{1}{16})$\\
\hline
$1,2$ & 0.14583 & --- --- & $(\frac{1}{12},\frac{1}{16})$  \\
$1,2$ & 0.39583 & $\frac{1}{4}$ & $(\frac{1}{3},\frac{1}{16})$ \\
$1,2$ & 0.64581 & $-\frac{1}{2}$ & $(\frac{1}{12},\frac{9}{16})$
\end{tabular}
\caption{Conformal dimensions of primary fields of the anti-ferromagnetic $\mathbb{Z}_3$ clock model with odd number of twist defects.
$m$ above is a label for the mixed sector Hamiltonian $\widetilde{H}^{(m)}$, or the $Q=e^{2 m \pi i/k}$ charge sector of $H^{(0)}$.
$\Delta P$ is the difference of the ``momentum" quantum numbers between any excited eigenstate and
the lowest energy eigenstate in the same charge sector and of the same system size.
The results shown above are obtained using ED for $N=6,8,10,12,14$.}
\label{Scaled_K3_AF_odd}
\end{table}

Similar rules apply  when $k\geq 5$. For $\mathbb{Z}_k$ clock model, the critical point is described by $U(1)_k$ CFT. In Table \ref{Scaled_K5_even} and \ref{Scaled_K6_even}, we present the numerical results of conformal dimensions for $k=5,6$ with both ferromagnetic and antiferromagnetic coupling in even number twist defect chain. In all of these cases, $h$ and $\bar{h}$ are consistent with the result for $U(1)_k$ CFT. Moreover, they always come in pairs, suggesting the partition function takes a diagonal form.

When $k=6$, if we consider odd-defect chain (Table \ref{Scaled_K6_F_odd}), for both ferromagnetic and antiferromagnetic coupling, the lowest several excitations
have $E_R=1/16+r^2/4k$, where $1/16$ is coming from the $\mathbb{Z}_2$ twist field operator and $r^2/24$ corresponds to the excitation in the original $U(1)_6$ CFT.
When $k=5$, for the odd-defect chain, if the coupling is antiferromagnetic, as shown in Table \ref{Scaled_K5_AF_odd}, the lowest several excitations  are still equal to $1/16+r^2/4k$.
For the ferromagnetic coupling, the quality of the numerical result is not fine enough due to strong finite size effect  and we cannot extract meaningful $h$ and $\bar{h}$. Nevertheless, these results suggest that for $k\neq 4$, the underlying CFT for odd-defect chain and even-defect chain are related through $\mathbb{Z}_2$ orbifolding and the extra twist defect in the odd-chain effectively introduces a $\mathbb{Z}_2$ twist field in the original CFT.


\begin{table}[htbp]
\centering
\begin{tabular}{c|c|c|c}
$m$ & $E_R$ & $\frac{L-1/2}{2\pi}\Delta P$ & $(h+n + \bar{h}+\bar{n})$ \\\hline
$0$ & 0.0625 & --- --- & $(0,\frac{1}{16})$ \\
$0$ & 0.563 & $-\frac{1}{2}$ & $(0, \frac{9}{16})$ \\
\hline
$1,4$ & 0.1125 & --- --- & $(\frac{1}{20},\frac{1}{16})$ \\
$1,4$ & 0.614 & $-\frac{1}{2}$ & $(\frac{1}{20},\frac{9}{16})$ \\
\hline
$2,3$ & 0.2628 & --- --- & $(\frac{1}{5},\frac{1}{16})$ \\
$2,3$ & 0.513 & $\frac{1}{4}$ & $(\frac{9}{20},\frac{1}{16})$
\end{tabular}
\caption{Conformal dimensions of primary fields of anti-ferromagnetic $\mathbb{Z}_5$ clock model with odd number of twist defects. $n$ above is a label for the mixed sector Hamiltonian $\widetilde{H}^{(m)}$, or the $Q=e^{2 m \pi i/k}$ charge sector of $H^{(0)}$.
$\Delta P$ is the difference of the ``momentum" quantum numbers between any excited eigenstate and
the lowest energy eigenstate in the same charge sector and of the same system size. The results shown above are obtained from ED results of $N=4,6,8,10,12$.
}\label{Scaled_K5_AF_odd}
\end{table}

\begin{table}[htbp]
\centering
\begin{tabular}{c|c|c|c}
$m$ & $E_R$ & $\frac{L-1/2}{2\pi}\Delta P$ & $(h+n + \bar{h}+\bar{n})$ \\\hline
$0$ & 0.0625 & --- --- & $(0,\frac{1}{16})$ \\
\hline
$1,5$ & 0.1042 & --- --- & $(\frac{1}{24},\frac{1}{16})$ \\
$1,5$ & 0.6047 & $-\frac{1}{2}$ & $(\frac{1}{24},\frac{9}{16})$ \\
\hline
$2,4$ & 0.2295 & --- --- & $(\frac{1}{6}, \frac{1}{16})$ \\
\hline
$3$ & 0.438 & --- --- & $(\frac{3}{8}, \frac{1}{16})$
\end{tabular}
\caption{Conformal dimensions of primary fields of ferromagnetic/anti-ferromagnetic $\mathbb{Z}_6$ clock model with odd number of twist defects.
$n$ above is a label for the mixed sector Hamiltonian $\widetilde{H}^{(m)}$, or the $Q=e^{2 m \pi i/k}$ charge sector of $H^{(0)}$.
$\Delta P$ is the difference of the ``momentum" quantum numbers between any eigenstate and
the lowest energy eigenstate in the same charge sector and of the same system size.
The results shown above are obtained from ED results of $N=4,5,6,7,8,9,10$.
}\label{Scaled_K6_F_odd}
\end{table}

\subsection{ $k=4$}
The $k=4$ odd-chain is much more complicated than $k>5$ cases. This is because when the chain consists of even number of defects, the $\mathbb{Z}_4$ clock model is already described by the $U(1)_2/\mathbb{Z}_2$ orbifold CFT, which is equivalent to the Ising$^2$ CFT.\cite{Ginsparg88,Teo2015} As shown in Table~\ref{Scaled_K4_even}, there are already some excitations with conformal dimension $h=\bar{h}=1/16$.

For the odd number twist defect chain, we find that this model cannot be described by further orbifolding
$\mathbb{Z}_2$ symmetry. We list the rescaled excitation energy $E_R$ shown in the second column of Table~\ref{Scaled_K4_odd}. Notice that the ground state has $E_R=1/16+1/64$, where the new $1/64$ excitation is  smaller than $h=1/32$ of the twist field in $\mbox{Ising}^2/\mathbb{Z}_2$ CFT.\cite{KLEMM1990} This suggests that the odd-defect chain cannot be described by the  $\mathbb{Z}_2$ orbifold CFT like other $k\neq 4$ cases. We further observe that for the excitations in  $m=0,2$ sectors, apart from $1/16$ part, the rest part of $E_R$ fits well with $t^2/64+n$ or $t^2/64+n+1/2$, where $t=1, 3, 5, 7$. Surprisingly, this $1/64$ excitation also shows up in the $SU(2)_1/D_4$ CFT ($D_4=\mathbb{Z}_4\rtimes\mathbb{Z}_2$ is the dihedral group at order 8) and is the conformal dimension for the four-fold symmetry sector.\cite{Teo2015} This coincidence motivates us to propose that the odd defect chain might be related with some $\mathbb{Z}_4$ orbifold CFT.

Here we briefly explain the physics in $SU(2)_1/D_4$ CFT and its connection with four-state Potts CFT. The self-dual Ashkin-Teller quantum chain model, in terms of $\mathbb{Z}_4$ clock variable, has the following Hamiltonian,\cite{Kohmoto1981}
\begin{align}
H=&-\sum_i\left[ \sigma_i+\sigma_i^{\dag}+\lambda (\sigma_i)^2\right. \nonumber\\
&\left. +\tau_i\tau_{i+1}^{\dag}+\tau_i^{\dag}\tau_{i+1}+\lambda(\tau_i)^2(\tau_{i+1})^2 \right]~.
\end{align}
For this model, as we vary $\lambda$ from 0 to 1, the model changes from $\mathbb{Z}_4$ clock model to four-state Potts model and remains critical for the whole regime for $\lambda$ between 0 and 1. This critical line is the famous Ashkin-Teller line and can be described by $U(1)/\mathbb{Z}_2$ orbifold CFT, where the compactification radius of $U(1)$ CFT changes as we vary $\lambda$.\cite{YANG1987}

For four-state Potts CFT, it corresponds to $U(1)_4/\mathbb{Z}_2$ CFT, which is also equivalent to $SU(2)_1/D_2$ CFT, where $D_2$ is the dihedral group at order 4 and is the double-cover of the $180^{\circ}$ rotations about the $x$, $y$, $z$-axes.\cite{Ginsparg88,DijkgraafVafaVerlindeVerlinde99} Actually, starting from the $SU(2)_1$ CFT,  we can get a family of orbifold CFTs by modding out the subgroup of $SU(2)$ (or called ADE classification).\cite{Ginsparg88,DijkgraafVafaVerlindeVerlinde99,CappelliAppollonio02,Teo2015} For $SU(2)_1/D_2$ CFT, it lies in the middle of this interesting series orbifold CFTs and has eleven characters which are reorganized in Table (\ref{u(1)_4/z_2}) in a more symmetric way. There exists a $S_3=\mathbb{Z}_3\rtimes\mathbb{Z}_2$ symmetry for $SU(2)_1/D_2$ CFT, which shuffles the twist fields $J_a$, $\sigma_a$ and $\tau_a$ ($a=1,2,3$) separately. In principle, we can orbifold the full $S_3$ symmetry and obtain $SU(2)_1/O$ CFT, where $O$ represents the octahedral group.\cite{Ginsparg88,DijkgraafVafaVerlindeVerlinde99,CappelliAppollonio02,Teo2015} However for our purpose in this paper, we only need to orbifold the two-fold symmetry and we obtain $SU(2)_1/D_4$ CFT which is equivalent to $U(1)_4/\mathbb{Z}_4$ CFT. This CFT has eight primary fields from the four-fold symmetry sector with conformal dimension $h=1/64+s(2s-1)/8$ or $h=33/64+s(2s-1)/8$ with $s=0,1,2,3$.\cite{Teo2015} These values actually are the same as $t^2/64$ or $t^2/64+1/2$ for $m=0,~2$ sectors in the second column of  Table~\ref{Scaled_K4_odd}.

We also compute the momentum by diagonalizing $\tilde{\mathcal{T}}$ operator and show $\Delta P$ in the third column of  Table~\ref{Scaled_K4_odd}. Based on $E_R$ and $\Delta P$, we list one possible decomposition $(h+n,\bar{h}+\bar{n})$ in the fourth column of Table~\ref{Scaled_K4_odd}. We compare this $\#/64$ in $h$ or $\bar{h}$ in $m=0,~2$ sectors with conformal dimension for primary fields in four-fold symmetry sector in $SU(2)_1/D_4$ CFT
and we find that they can partially match up. Moreover, $SU(2)_1/D_4$ CFT also has two-fold symmetry sector corresponding to twofold rotation about a diagonal axis like (110) which actually has $h=1/16,~9/16$ and is consistent with $m=1,~3$ sectors.  At this moment, it is unclear why there is connection between odd number twist defect chain and $SU(2)_1/D_4$ ($U(1)_4/\mathbb{Z}_4$) CFT.
What is puzzling is that the even chain and odd chain are not seemingly related by orbifold. The even chain has an Ising$^2$ CFT with the compactification radius $R=1$, but the odd chain is suggestively $SU(2)_1/D_4$ which has a larger radius $R=\sqrt{2}$.
We leave this disagreement for future studies.



\begin{table}[htbp]
\centering
\begin{tabular}{c|c|c|c|c|c}
$\chi$  & $\chi_I$ & $\chi_J^a$ & $\chi_{\phi} $ & $\chi^a_{\sigma}$ & $\chi^a_{\tau}$ \\\hline
$d_\chi$ & $1$ & $1$ & $2$ & $2$  &  $2$ \\
$h_{\chi}$ & $0$ & $1$ & $\frac{1}{4}$ & $\frac{1}{16}$ & $\frac{9}{16}$
\end{tabular}
\caption{The quantum dimensions $d_\chi$, conformal dimension $h_{\chi}$ of characters for chiral $U(1)_4/\mathbb{Z}_2$ CFT, where $a=1,2,3$.
}\label{u(1)_4/z_2}
\end{table}


\begin{table}[htbp]
\centering
\begin{tabular}{c|c|c|c}
$m$ & $E_R$ & $\frac{L-1/2}{2\pi}\Delta P$ & $(h+n, \bar{h}+\bar{n})$ \\\hline
$0$ & $0.07812=\frac{1}{16}+\frac{1}{64}$ & --- --- & $(\frac{1}{64},\frac{1}{16})$  \\
$0$ & $0.70312=\frac{9}{16}+\frac{9}{64}$ & $-\frac{1}{2}$ & $(\frac{5}{64},\frac{5}{8})$\\
$0$ & $0.82812=\frac{1}{16}+\frac{49}{64}$ & $\frac{1}{2}$ & $(\frac{41}{64},\frac{3}{16})$ \\
$0$ & $0.95312=\frac{9}{16}+\frac{25}{64}$ & $1$ & $(\frac{61}{64},0)$\\
$0$ & $1.07812=\frac{17}{16}+\frac{1}{64}$ & $-1$ & $(\frac{1}{64},\frac{17}{16})$\\
$0$ & $1.07812=\frac{17}{16}+\frac{1}{64}$ & $1$ & $(\frac{65}{64},\frac{1}{16})$\\
$0$ & $1.32812=\frac{9}{16}+\frac{49}{64}$ & $-\frac{1}{2}$ & $(\frac{25}{64},\frac{15}{16})$\\
\hline
$1,3$ & 0.12500 & --- --- & $(\frac{1}{16},\frac{1}{16})$\\
$1,3$ & 0.62500 & $-\frac{1}{2}$ & $(\frac{1}{16},\frac{9}{16})$\\
$1,3$ & 0.62500 & $\frac{1}{2}$ & $(\frac{9}{16},\frac{1}{16})$\\
$1,3$ & 1.12500 & $-1$ & $(\frac{1}{16},\frac{17}{16})$\\
$1,3$ & 1.12500 & $1$ & $(\frac{17}{16},\frac{1}{16})$\\
$1,3$ & 1.12500 & $0$ & $(\frac{9}{16},\frac{9}{16})$\\
$1,3$ & 1.62495 & $\frac{3}{2}$ & $(\frac{25}{16},\frac{1}{16})$\\
$1,3$ & 1.62495 & $-\frac{3}{2}$ & $(\frac{1}{16},\frac{25}{16})$\\
\hline
$2$ & $0.20312=\frac{1}{16}+\frac{9}{64}$ & --- --- & $(\frac{3}{16},\frac{1}{64})$\\
$2$ & $0.45312=\frac{1}{16}+\frac{25}{64}$ & $-\frac{1}{2}$ & $(\frac{1}{16},\frac{25}{64})$\\
$2$ & $0.57812=\frac{9}{16}+\frac{1}{64}$ & $0$ & $(\frac{3}{8},\frac{13}{64})$\\
$2$ & $1.20312=\frac{17}{16}+\frac{9}{64}$ & $1$ & $(\frac{19}{16},\frac{1}{64})$\\
$2$ & $1.20312=\frac{17}{16}+\frac{9}{64}$ & $-1$ & $(\frac{3}{16},\frac{65}{64})$\\
$2$ & $1.32812=\frac{9}{16}+\frac{49}{64}$ & $-\frac{3}{2}$ & $(0,\frac{85}{64})$\\
$2$ & $1.45309=\frac{17}{16}+\frac{25}{64}$ & $-\frac{3}{2}$ & $(\frac{1}{16},\frac{89}{64})$
\end{tabular}
\caption{Hypothetical conformal dimensions $(h+n, \bar{h}+\bar{n})$ of primary fields of the ferromagnetic/antiferromagnetic $\mathbb{Z}_4$ clock model with odd number of twist defects. $n$ above is a label for the mixed sector Hamiltonian $\widetilde{H}^{(m)}$, or the $Q=e^{2 m \pi i/k}$ charge sector of $H^{(0)}$.
$\Delta P$ is the difference of the ``momentum" quantum numbers between any excited eigenstate and
the lowest energy eigenstate in the same charge sector and of the same system size.
The results shown above are obtained from ED results of $N=6,7,8,9,10,11,12,13$.
}\label{Scaled_K4_odd}
\end{table}

\section{Conclusion}
\label{conclusion}
In this first part of the paper, we study the twofold twist defect chain at critical point. We demonstrate that for even number of twist defects, it maps to the $\mathbb{Z}_k$ clock model with periodic boundary condition (up to some phase factor), while for the odd number case, it is equivalent to the $\mathbb{Z}_k$ clock model with a duality twisted boundary condition. The translation symmetry in the twist defect chain model becomes the Kramers-Wannier duality symmetry in the $\mathbb{Z}_k$ clock model. This symmetry operator can be generated by a series of braiding operators for twist defects, and is discussed in section~\ref{sec:trans}.


In the second part of the paper, we numerically investigate the defect chain model at its self-dual critical point. We first extract the conformal dimensions for the primary fields in the even-defect chain model and find that they match up with that of the $\mathbb{Z}_k$ clock CFT. We then turn to study the underlying CFT for odd-defect chains and we observe the energy spectrum is shifted, where the energy difference is caused by the twist field in the orbifold CFT. We find that when $k\neq 4$, the odd-defect chain is described by orbifolding the $\mathbb{Z}_2$ symmetry in the even-defect chain CFT. On the other hand, when $k=4$, there is a mysterious $1/64$ excitation in the spectrum of odd-defect chain which turns out to be related with the $\mathbb{Z}_4$ twist field in the $SU(2)_1/D_4$ orbifold CFT. Our model can be generalized to twist defect with other symmetries and can be used to realize more complicated orbifold CFTs.

\begin{acknowledgements}
We acknowledge useful discussion with Bryan Clark, Eduardo Fradkin and Andreas Ludwig. XY was supported from the DOE through Grant No. SciDAC FG02-12ER46875. XC was supported by a postdoctoral fellowship from the the Gordon and Betty Moore Foundation, under the EPiQS initiative, Grant GBMF4304, at the Kavli Institute for Theoretical Physics. This work is supported by the NSF under Grant No.~DMR-1653535 (JCYT) and DMR-1408713 (XC). AR was supported by the German Research Foundation (DFG) through grants ZI 513/2-1 and
HE 7267/1-1.
\end{acknowledgements}

\appendix
\section{Braiding operators}
\label{app:braiding}
One can write out the $B$-operators using the braiding rules of Eq. \eqref{braiding_eq}. These $B$-operators, as shown pictorially in Fig.~\ref{b_even} and Fig.~\ref{b_odd} would generate the duality transformation and translation in the $\mathbb{Z}_k$ clock model variables.

For $k=3$,
\begin{equation}
B_{2j-1} = \frac{1}{\sqrt{3}}[(\sigma_j+\sigma^{\dagger}_j)+\omega]
\end{equation}
\begin{equation}
B_{2j} = \frac{1}{\sqrt{3}} [(\tau_j\tau^{\dagger}_{j+1}+\tau^{\dagger}_j\tau_{j+1})+\omega]
\end{equation}
For $k=4$,
\begin{equation}
B_{2j-1} = \frac{1}{\sqrt{4}} [(\sigma_j+\sigma^{\dagger}_j)+\omega^{\frac{3}{2}}\sigma_j^2+\omega^{-\frac{1}{2}}]
\end{equation}
\begin{equation}
B_{2j} = \frac{1}{\sqrt{4}} [(\tau_j\tau^{\dagger}_{j+1}+\tau^{\dagger}_j\tau_{j+1})+\omega^{\frac{3}{2}}\tau^2_j\tau^2_{j+1}+\omega^{-\frac{1}{2}}]
\end{equation}
For $k=5$,
\begin{equation}
B_{2j-1}  \frac{1}{\sqrt{5}} [(\sigma_j+\sigma^{\dagger}_j)+\omega^{4}(\sigma_j^2+\sigma^{\dagger 2}_j)+\omega^2]
\end{equation}
\begin{equation}
B_{2j} = \frac{1}{\sqrt{5}} [(\tau_j\tau^{\dagger}_{j+1}+\tau^{\dagger}_j\tau_{j+1})+\omega^{4}(\tau^2_j\tau^{\dagger 2}_{j+1}+\tau^{\dagger 2}_j\tau^2_{j+1})+\omega^2]
\end{equation}
For $k=6$,
\begin{equation}
B_{2j-1} = \frac{1}{\sqrt{6}} [(\sigma_j+h.c.)+\omega^{\frac{3}{2}}(\sigma_j^2+h.c.)+\omega^4\sigma^{3}_j+\omega^{-\frac{1}{2}}]
\end{equation}
\begin{equation}
B_{2j}  = \frac{1}{\sqrt{6}} [(\tau_j\tau^{\dagger}_{j+1}+h.c.)+\omega^{\frac{3}{2}}(\tau^2_j\tau^{\dagger 2}_{j+1}+h.c.)+\omega^4\tau^3_j\tau^3_{j+1}+\omega^{-\frac{1}{2}}]
\end{equation}
Notice all $B$-operators defined above satisfy $BB^{\dagger}=\mathbb{I}$, and commute with the charge operator $Q$.

\section{$U(1)_{k}/Z_2$ orbifold CFT}
\label{U_1_orb}
The  chiral $U(1)_{k}$ ($k\in \mathbb{Z}^+$) CFT describes a compact free bosonic field $\phi$ identified modulo $2\pi R$ with $R=\sqrt{2k}$. There are $2k$ primary fields $V_r=e^{ir\phi/\sqrt{2k}}$, which are vertex operators and satisfy the $\mathbb{Z}_{2k}$ abelian fusion rules $V_{[r]}\times V_{[r^{\prime}]}=V_{[r+r^{\prime}]}$ with $[r]$ defined as $r$ mod $2k$.  The corresponding characters are
\begin{align}
\nonumber\chi_{r}(\tau)=&\frac{1}{\eta(\tau)}\sum_n q^{k(n+\frac{r}{2k})^2}\\
=&\frac{1}{\eta(\tau)}\Theta_{r,2k}(q)
\label{theta_f}
\end{align}
where $r\in \mathbb{Z}$ and satisfies $0\leq r<2k$, $q=e^{i2\pi\tau}$ and $\eta(\tau)$ is the Dedekind eta function
\begin{align}
\eta(\tau)=q^{\frac{1}{24}}\prod_{n=1}^{\infty}(1-q^n)
\label{eta_func}
\end{align}

Under $\mathcal{T}$ transformation
\begin{equation}
\Theta_{r,2k}(\tau+1)=e^{2\pi i\frac{r^2}{4k}}\Theta_{r,2k}(\tau)
\end{equation}

Under $\mathcal{S}$ transformation
\begin{equation}
\Theta_{r,2k}(-\frac{1}{\tau})=\sqrt{\frac{-i\tau}{2k}}\sum_{n}\Theta_{r^{\prime},2k}(\tau)e^{-2\pi i\frac{r^{\prime}r}{2k}}
\end{equation}

This CFT is invariant under $\mathbb{Z}_2$ symmetry $\phi\leftrightarrow-\phi$, which corresponds to the charge conjugation symmetry for the vertex operators and exchanges $V_r$ and $V_{2k-r}$. After orbifolding $\mathbb{Z}_2$ symmetry, the model is projected to the $\mathbb{Z}_2$ invariant states with the twisted sectors also need to be included.\cite{DijkgraafVafaVerlindeVerlinde99} The $U(1)_{k}/\mathbb{Z}_2$ orbifold CFT has $k+7$ characters which can be constructed from the partition function of $\phi$ field with twisted boundary condition in time and spatial directions. The conformal dimension of the characters are listed in Table (\ref{u(1)_k}). Notice that four of them are coming from the twist field operators and have conformal dimensions equal to $\frac{1}{16}$ or $\frac{9}{16}$. These characters can be used to construct a modular invariant non-chiral partition function with a diagonal form partition function $Z=\sum_{i=1}^{k+7}|\chi_i|^2$.

\begin{table}[htbp]
\centering
\begin{tabular}{lll}
$\chi$ & $d_\chi$ & $h_{\chi}$  \\\hline
$\chi_{\mbox{I}}$ & $1$ & $0$  \\
$\chi_J$ & $1$ & $1$ \\
$\chi_{k}^l,\ l=(0,1)$ & $1$ & $\frac{k}{4}$ \\
$\chi_{r},\ r=(1,...,k-1)$ & $2$ & $\frac{r^2}{4k}$ \\
$\chi^l_{\sigma},\ l=(0,1)$ & $\sqrt{k}$ & $\frac{1}{16}$ \\
$\chi^l_{\tau},\ l=(0,1)$ & $\sqrt{k}$ & $\frac{9}{16}$ \\
\end{tabular}
\caption{The quantum dimensions $d_\chi$, conformal dimension $h_{\chi}$ of characters for chiral $U(1)_{k}/\mathbb{Z}_2$ CFT .
}\label{u(1)_k}
\end{table}

Here we list several well-known results for lattice model. For $U(1)_k/\mathbb{Z}_2$ CFT, when $k=2$, the orbifold CFT has nine primary fields and corresponds to $\mbox{Ising}^2$ CFT with nine primary fields. When $k=3$, it corresponds to the $Z_4$ parafermion CFT. When $k=4$, it is the four-state Potts CFT.

\section{Numerical method to compute the ground state energy $E_0$}
\label{gs_energy}
For these three parameters in $E_0=\epsilon_{\infty}L - \pi v c/6L$ defined in Eq.\eqref{cft_spec}, $\epsilon_{\infty}$ can be found using iDMRG to high accuracy.
The central charge $c$ can be obtained by fitting to the scaling form of the entanglement entropy.
Given the ground state of a $1+1$d critical chain of length $L$ with open boundary conditions,
if we consider a consecutive block of size $L_A$ starting from the left (or right) edge,
the von Neumann entanglement entropy of that block has been shown to exhibit the following scaling behavior \cite{CalabreseCardy04}
\begin{equation}
S(L_A) = \frac{c}{6}\log \left( \frac{L}{\pi} \sin\frac{\pi L_A}{L} \right) + S_0
\label{S_scaling}
\end{equation}
where $S_0$ is a constant piece. Numerically, we use DMRG to obtain the ground state wave function of
a finite chain with length $L$ and open boundary conditions, and determine the central charge $c$ by
fitting the numerical results of $S(L_A)$ onto Eq. \eqref{S_scaling}. Finally,
we can extract the sound velocity $v$ based on the ground state energy $E_0=\epsilon_{\infty}L - \pi v c/6L + O(L^{-2})$
(obtained by ED). Polynomial extrapolation in terms of $1/L$ is used to mitigate the finite size correction ($ O(L^{-2})$).

\begin{figure}[h]
	\centering
	\includegraphics{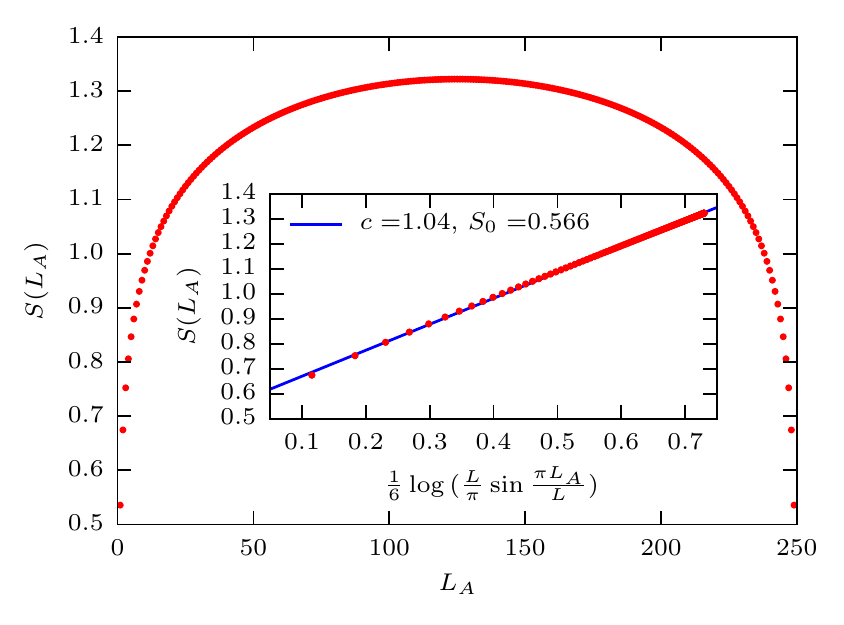}
	\caption{Entanglement entropy $S$ vs. subsystem size $L_A$, for a open boundary $\mathbb{Z}_4$ critical chain with $L=250$.
  The central charge $c=1$ can be read off from the linear fit in the inset, which is based on Eq. \eqref{S_scaling}.}
	\label{cc_K4}
\end{figure}

\begin{figure}[h]
	\centering
	\includegraphics{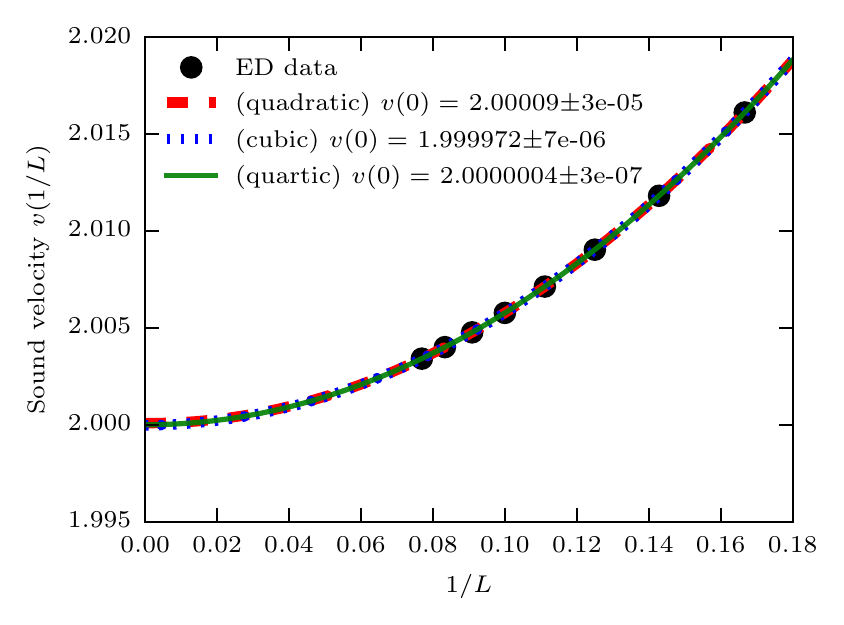}
	\caption{Polynomial extrapolation of the sound velocity $v$ for finite periodic critical $\mathbb{Z}_4$ clock model,
  based on $E_0=\epsilon_{\infty}L - \pi v c/6L + O(L^{-2})$. The extrapolated value is very close to the exact value of 2.}
	\label{v_K4}
\end{figure}

As an example, in the following we demonstrate how we obtain $\epsilon_{\infty}$, $c$, and $v$ for the critical $\mathbb{Z}_4$ clock model.
$k=4$ is a special case where the anti-ferromagnetic and ferromagnetic critical clock models are related by using $\sigma\tau =\omega \tau\sigma$.
First of all, through iDMRG we find $\epsilon_{\infty} =2.54647904$, which agrees with the exact value of $−8/\pi$  up to
 very high precision.\cite{Gehlen1987} Secondly, DMRG calculations, as can be see in Fig. \ref{cc_K4}, show that c = 1.
 Then, using the ED calculations for lengths $L = 6,7,8,9,10,11,12,13$, we calculate the sound velocity,
 and extrapolate it to $v = 2.0000$ in the $1/L\to 0$ limit as in Fig. \ref{v_K4}, compared to the exact value of $v = 2$. \cite{Gehlen1987} All the parameters above are obtained
 to high accuracy.

The parameters of $\epsilon_{\infty}$, $c$ and $v$ for $\mathbb{Z}_3$, $\mathbb{Z}_4$, $\mathbb{Z}_5$, and $\mathbb{Z}_6$ clock models
are listed in Table~\ref{ZkData}. In addition to the comparison between the numerical and exact results for the $k = 4$ case,
our results for the ferromagnetic and antiferromagnetic $k=3$ cases also match up  extremely well with previous numerical results \cite{Li2015},
and the Bethe ansatz solutions \cite{Albertini1992397} of $\epsilon_{\infty}=-2\sqrt{3}/\pi-4/3$, $v=3\sqrt{3}/2$ (ferromagnetic),
and  $\epsilon_{\infty}=-\sqrt{3}/\pi-3\sqrt{3}/2+4/3$, $v=3\sqrt{3}/4$ (anti-ferromagnetic).

\section{Mitigating finite size effects with polynomial extrapolation}
\label{rescaled_energy}
As in Eq. \eqref{even_rescaled_E} and \eqref{odd_rescaled_E}, the conformal dimensions calculated using exact diagonalization are
accompanied by finite size corrections that are of $O(1/L)$. To mitigate these finite size effects, we extrapolate the
conformal dimensions at different system sizes using polynomials of various orders of $1/L$, and check whether consistent results
can be obtained in the limit of $1/L \to 0$. Reliable results are then listed in the tables.

\begin{figure}[h]
	\centering
	\includegraphics{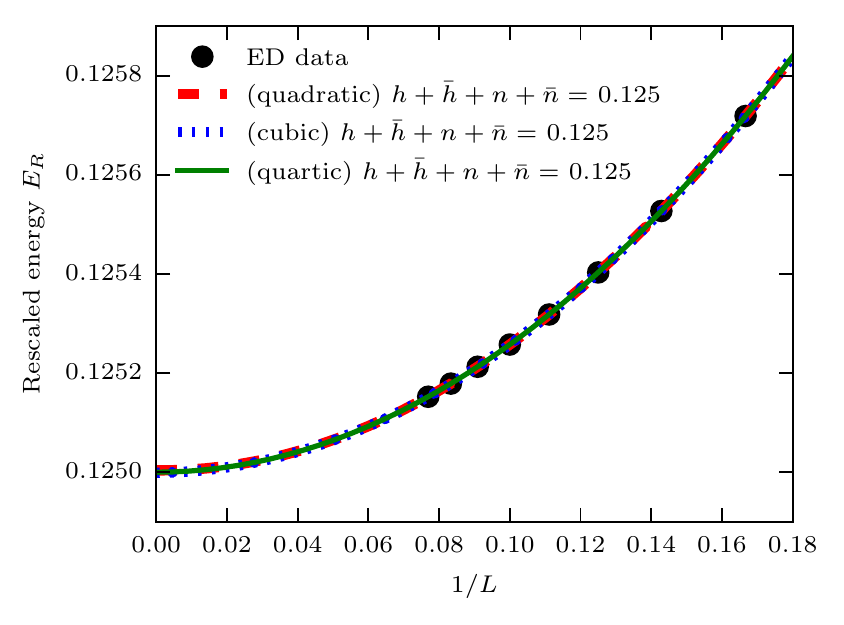}
	\caption{Polynomial extrapolation of the first excited state's rescaled energy for the critical $\mathbb{Z}_4$ clock model
  (even-defect chain). This is listed in the $q=1,3$ sector of Table. \ref{Scaled_K4_even}.}
  \label{first_Z4}
\end{figure}

\begin{figure}[h]
	\centering
	\includegraphics{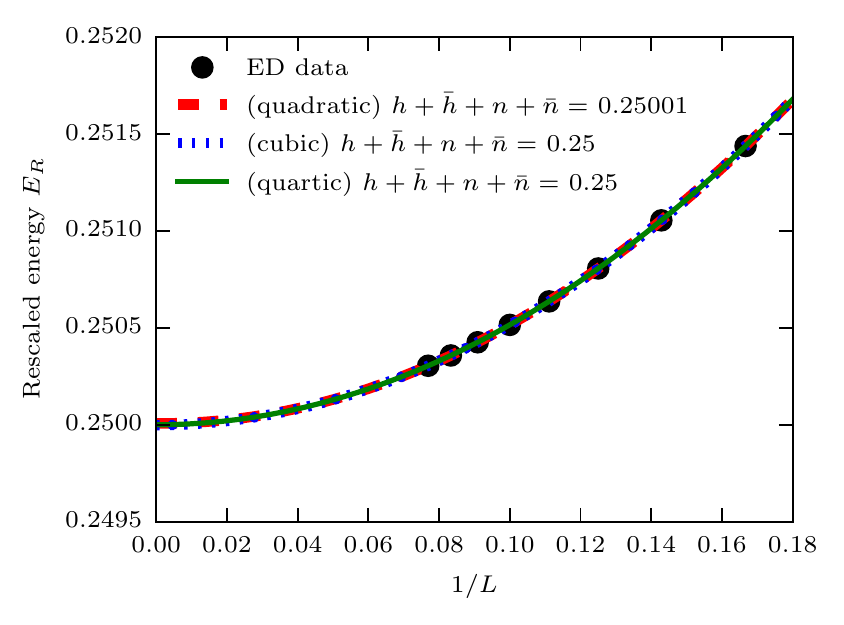}
	\caption{Polynomial extrapolation of the second excited state's rescaled energy for the critical $\mathbb{Z}_4$ clock model
  (even-defect chain). This is listed in the $q=2$ sector of Table. \ref{Scaled_K4_even}.}
  \label{second_Z4}
\end{figure}

\begin{figure}[h]
	\centering
	\includegraphics{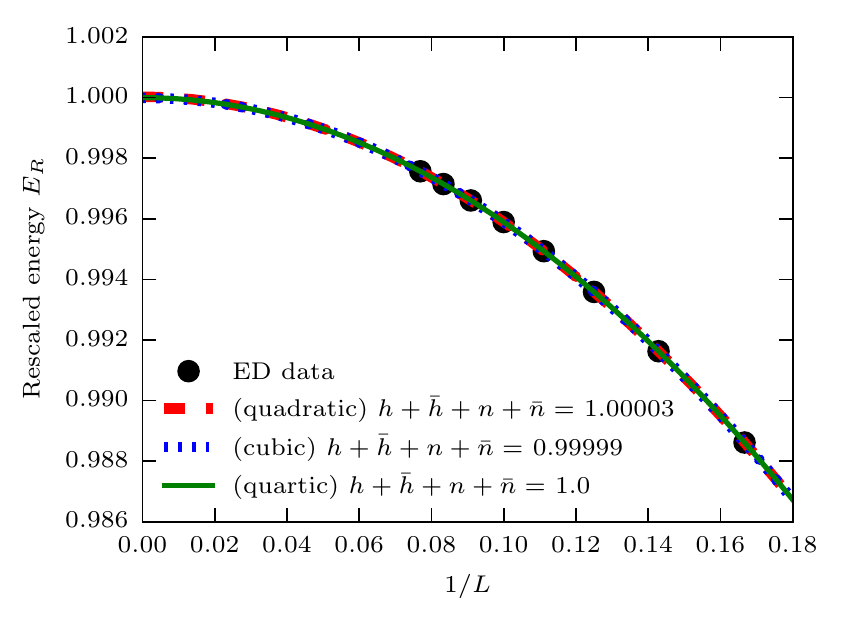}
	\caption{Polynomial extrapolation of the third excited state's rescaled energy for for the critical $\mathbb{Z}_4$ clock model
  (even-defect chain). This is listed in the $q=0,~ 2$ sector of Table. \ref{Scaled_K4_even}.}
  \label{third_Z4}
\end{figure}

\begin{figure}[h]
	\centering
	\includegraphics{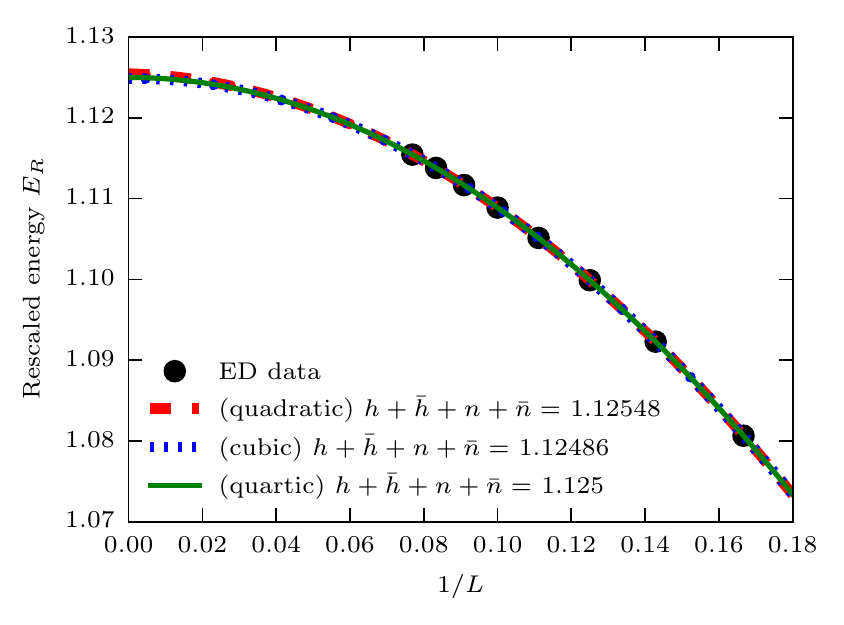}
	\caption{Polynomial extrapolation of the fifth excited state's rescaled energy for  the critical $\mathbb{Z}_4$ clock model
  (even-defect chain). This is listed in the $q=1,~3$ sector of Table. \ref{Scaled_K4_even}.}
  \label{fifth_Z4}
\end{figure}

\begin{figure}[h]
	\centering
	\includegraphics{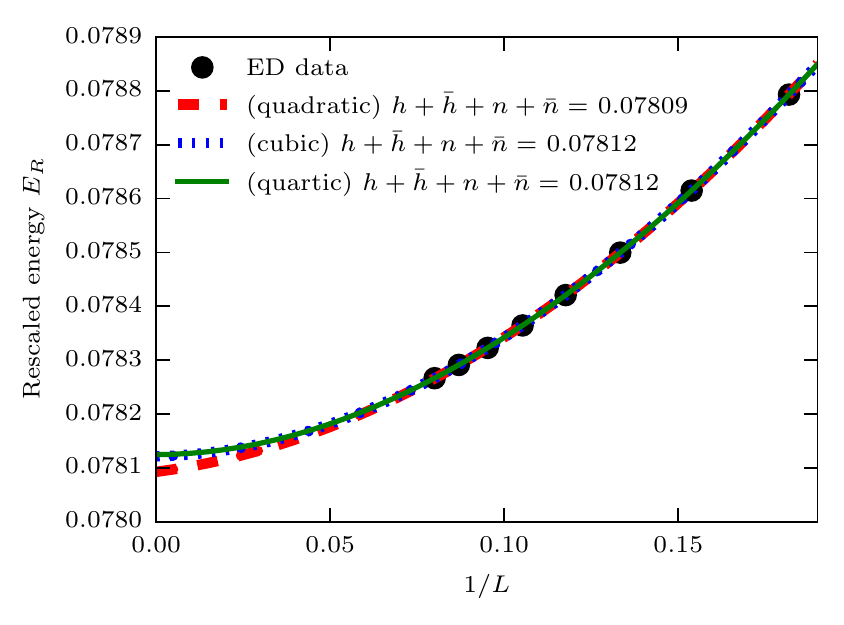}
	\caption{Polynomial extrapolation of the ground state's rescaled energy for $\mathbb{Z}_4$ odd-defect chain. This is listed in the $q=0$ sector of Table. \ref{Scaled_K4_odd}.}
	\label{gs_odd_Z4}
\end{figure}

\begin{figure}[h]
	\centering
	\includegraphics{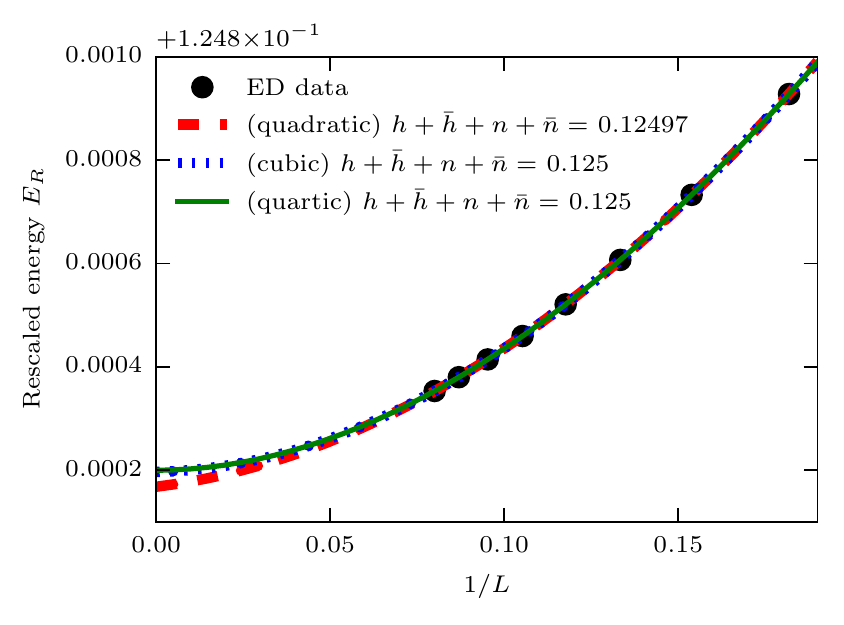}
	\caption{Polynomial extrapolation of the first excited state's rescaled energy for $\mathbb{Z}_4$ odd-defect chain. This is listed in the $q=1,3$ sector of Table. \ref{Scaled_K4_odd}.}
	\label{first_odd_Z4}
\end{figure}

\begin{figure}[h]
	\centering
	\includegraphics{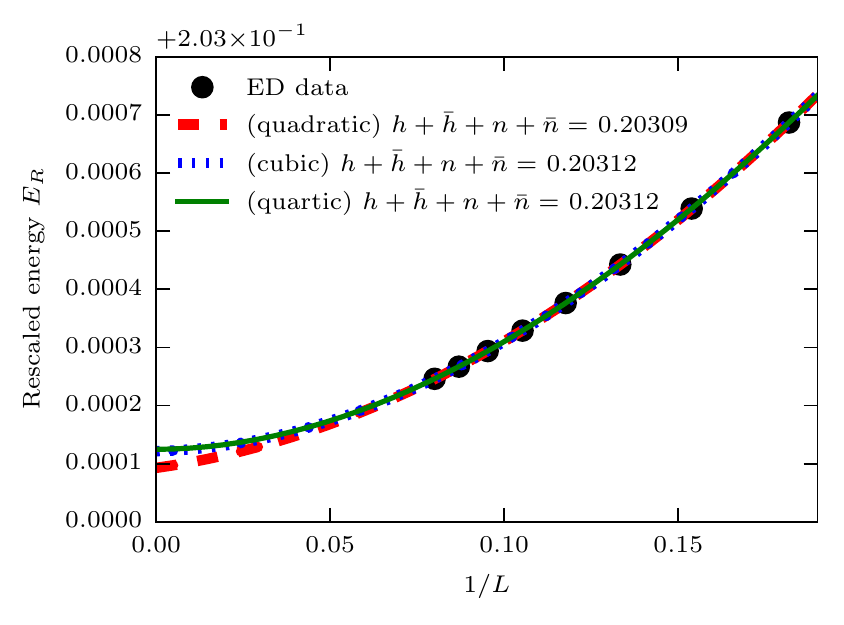}
	\caption{Polynomial extrapolation of the third excited state's rescaled energy for $\mathbb{Z}_4$ odd-defect chain. This is listed in the $q=2$ sector of Table. \ref{Scaled_K4_odd}.}
	\label{third_odd_Z4}
\end{figure}

\begin{figure}[h]
	\centering
	\includegraphics{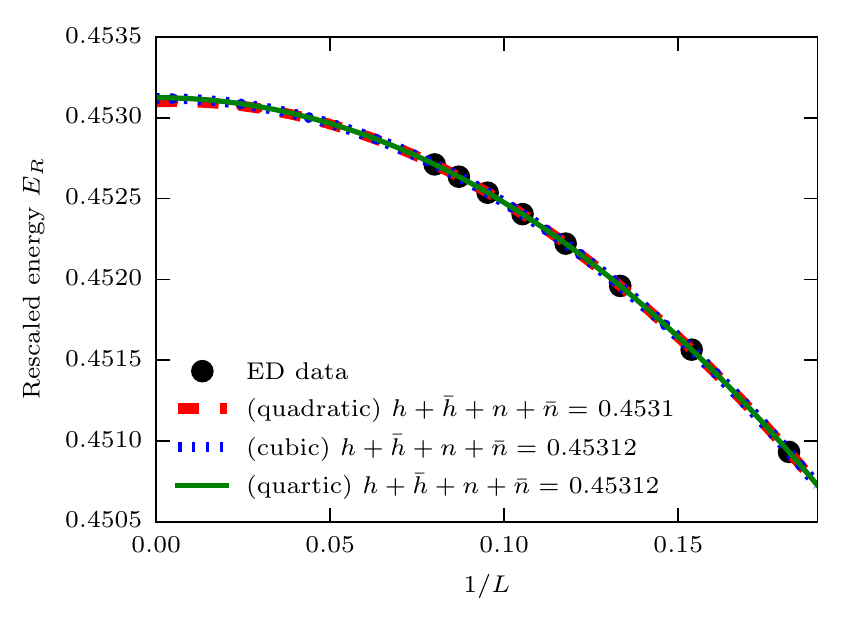}
	\caption{Polynomial extrapolation of the fourth excited state's rescaled energy for $\mathbb{Z}_4$ odd-defect chain. This is listed in the $q=2$ sector of Table. \ref{Scaled_K4_odd}.}
	\label{fourth_odd_Z4}
\end{figure}

In particular, we show several figures (Fig.~\ref{first_Z4}, Fig.~\ref{second_Z4}, Fig.~\ref{third_Z4}, Fig.~\ref{fifth_Z4}, Fig.~\ref{gs_odd_Z4}, Fig.~\ref{first_odd_Z4},  Fig.~\ref{third_odd_Z4}, Fig.~\ref{fourth_odd_Z4}) for $\mathbb{Z}_4$ clock models below to illustrate this idea.


\bibliography{chain_reference}

\end{document}